\documentclass[11pt]{article}

\newcommand{\lbl}[1]{\label{#1}}
\newcommand{\bib}[1]{\bibitem{#1}}

\usepackage{amsxtra}
\usepackage{amsmath}

\usepackage{graphicx}
\graphicspath{%
    {converted_graphics/}
    {C:/COSMOS/Integration/Physruler/}
    {./}
    {/}
    {markc(t)/}
    {Z3/}
}

\begin{document}

\newtheorem{thm}{Theorem}
\newtheorem{lm}{Lemma}
\newtheorem{prop}{Proposition}

\title{On a c(t)-Modified Friedman-Lemaitre-Robertson-Walker Universe}
 \author{Robert C. Fletcher\footnote{Mailing address: 1000 Oak Hills Way, Salt Lake City, UT 84108.  robert.c.fletcher@utah.edu}
\\
{\small Bell Telephone Laboratories (ret)}\\
{\small Murray Hill, New Jersey.}}  
\date{}

\maketitle

\begin{quote} {\small \begin{center} {\bf Abstract}\end{center}

This paper presents a compelling argument for the physical light speed in the homogeneous and isotropic Friedman-Lemaitre-Robertson-Walker (FLRW) universe to vary with the cosmic time coordinate t of FLRW.   
It will be variable when the radial co-moving differential coordinate of FLRW 
is interpreted as physical and therefor transformable by a Lorentz transform locally to differentials of stationary physical coordinates.  Because the FLRW differential radial distance has a time varying coefficient a(t), 
in the limit of a zero radial distance the light speed c(t) becomes  time varying,  proportional to the square root of the derivative of a(t)

Since we assume homogeneity of space, this derived c(t) is the physical light speed for all events in the FLRW universe.  This impacts the interpretation of astronomical observations of distant phenomena that are sensitive to light speed.  
 
A transform from FLRW 
is shown to have a physical radius out to all radial events in the visible universe.  This shows a finite horizon beyond which there are no galaxies and no space.

The general relativity (GR) field equation to determine $a(t)$ and $c(t)$ is maintained by using a variable gravitational constant and rest mass that keeps constant the gravitational and particle rest energies. This keeps constant the proportionality constant between the GR tensors of the field equation and conserves the stress-energy tensor of the ideal fluid used in the FLRW GR field equation.

In the same way all of special and general relativity can be extended to include a variable light speed.\\\

}\end{quote}

keywords: cosmology theory---variable light speed---elementary particles---distances and red shifts---physical constants---relativity \\\

\tableofcontents

\section{ Introduction} \lbl{intro}

Here I will use a significantly different approach than other attempts in the literature to investigate a variable speed of light.  Those mostly tried to find a new cosmology to provide alternatives to inflation in order to resolve horizon and flatness problems\cite{Bas}\cite{HSK}\cite{Mag1}\cite{Mag3}\cite{EU}.  In common with those approaches, the present approach is a major departure from the prevailing paradigm that the speed of light is constant.  However, my calculation of a variable light speed $c(t)$ seems to be consistent with being interpreted as physical in the FLRW universe, but is a variable function of the Chronometric Invariant observable constant light speed $c$, dependent on the specific conditions in this universe compared with the more general universes treated by A. Zelmanov\cite{Z}.

I derive a variable light speed using the same assumptions used for almost a century, except that I allow for a variable light speed: (1) that light speed (even though variable) is independent of the velocity of the observers, (2) that the universe is homogeneous and isotropic, and (3) that the radial FLRW differential variables derivable for this universe represent physical time and distance.  
The first assumption leads to a Lorentz transform between moving observers, extended to allow a variable light speed; the second leads to the FLRW metric\cite{R}\cite{Wa}\cite{We1} that allows for a variable light speed; and the third allows us to locally apply the extended Lorentz transform from the FLRW time $dt$ and radial distance 
$a(t)d\chi$ to the stationary time $dT$ and distance $dR$.  We show that this requires a variable physical light speed to be $c(t) \propto \sqrt{\frac{da}{dt}}$ in order to be consistent with the time varying distance differential of FLRW. 
 This is done by expanding the physical time and distance along a stationary rod 
in a power series of the FLRW co-moving coordinate $\chi$   
 and extrapolating to zero $\chi$.  We assume 
(fourth assumption) that the Lorentz transform remains valid 
from the origin out to at least the lowest power $\chi$ and therefor the lowest power of the velocity between the two frames.
This derivation is fairly simple and covers only the first 11 pages of this document.  
The remainder of the document addresses the reasonableness and implications of this derivation.

We find two different systems of full radial transformed coordinates from FLRW, good for all distances, whose differentials close to the origin have a 
Minkowski metric.  These transforms all have the same variable light speed at the origin as the power series expansion, a universality that I find persuasive. 

For a homogeneous universe, since the origin can be placed on any galactic point, this means that this variable physical light speed enters all our physical laws throughout the universe.  In particular the standard candles like the Supernovae Ia\cite{P'}\cite{R1}\cite{RInv}\cite{T} and galactic clusters\cite{A} can be affected by the higher light speeds present for distant clusters.


 To maintain unchanged the field equation of general relativity, we assume the gravitational ``constant'' $G$ to be time varying, but keep constant 
the proportionality function between the GR tensors of the field equation.  This is done by assuming the particle rest energy and the Newton gravitational energy to be constant.  This also conserves the stress energy tensor of an ideal fluid used in the GR field equation for an FLRW universe. 

We can express the gravitational field in transformed stationary coordinates using Riemannian geometry.  In the region near the origin for a flat universe this field increases linearly with distance just like the Newtonian field for a spherical distribution of uniform mass density.  

A surprise bonus from this endeavor is that one of the radial transforms has a physical distance to all parts of the universe.  Even though three rigid accelerated axes are inadequate to describe three dimensional motion, it is apparently possible to find one rigid axis to measure radial distance, at least for a homogeneous FLRW universe, although the transformed time on this axis becomes non-physical at large distances.  This shows that in the coordinates of the rigid frame attached to the origin that the universe is contained within an expanding spherical shell outside of which there are no galactic points and no space.

I also outline in the Appendix how not only Lorentz, but all of the vectors and tensors of Special  Relativity can be extended to include a variable light speed so they can be used in the standard field equation of General Relativity.\\\

\section{The derivation of c(t)}

\subsection{Assumptions} \lbl{ass}

Only four assumptions are needed for the derivation of $c(t)$.  The first three are the same assumptions for special relativity and for the universe that have been made for almost a century.  What is new is the allowance for the possibility that the physical light speed is variable.  We will use ``line element'' to describe the invariant $ds$ and ``metric'' to describe the particular differential coordinates that equal $ds$.  We will be considering only radial motion in a spherically symmetric universe.
\\

$i$.  ``The physical light speed is the same for all co-located observers who may be moving at various velocities in an accelerating field.''\\

From this we can derive the extended Lorentz transform 
($\hat{L}$) between such observers, even when the light speed and velocities are variable (App \ref{LT}).  Each observer will have an extended Minkowski metric ($\hat{M}$). \\

$ii$. ``The universe is isotropic and homogeneous in space.''\\

From this we can derive an extended FLRW metric (App \ref{FLRW}) that allows for a variable light speed $c(t)$, where $t$ is the physical time on the co-moving galactic points of the FLRW solution.  This derivation depends only on the assumed symmetry and not on the general relativistic field equation. 
$\chi$ is a co-moving radial coordinate with which a galactic point (representing a galaxy) stays constant.  $a(t)$ is a universe scale factor that multiplies $d\chi$ in the metric.  \\ 

Definitions: A ``rigid frame'' is a plane where at any given time none of its spatial points has movement compared to any other spatial point. 
We will call ``physical'' those coordinate whose differentials over some interval of time or distance are the same at any two inertial points along a rigid frame connecting them .
 Moving rigid frames will generally have different physical coordinates from each other, although all will use the same invariant units when representing clock and ruler readings.  We call these coordinates physical because they represent those physical clocks and rulers on an inertial rigid frame on earth whose units are invariant. For instance, the physical clock might be the spectral frequency $\nu _s$ of a  standard atom (assumed invariant) and the physical ruler be $c_0/\nu _s$.  Physical differentials on any moving frame will be related by an invariant Minkowski metric $\hat{M}$  (see Appendix \ref{LT}).  Any other coordinate that represents the reading on one of these physical instruments we will also call physical  (see assumption v).
  Physical velocity of a moving point is the ratio of the differential physical distance that the point moves in a differential physical time when both time and distance coordinates are at the same location as the moving point.  
\\

$iii$.``The FLRW time and radial differentials $dt$ and $a(t)d\chi$ are differentials of physical coordinates.'' \\

This is a usual assumption. It is reasonable since the radial motion of the FLRW metric is $\hat{M}$ in these differentials.  Thus, galactic points are on rigid inertial frames.  With this assumption the radial physical light speed is $a (\frac{\partial \chi}{\partial t})_s$, and the physical radial velocity $V$ of a moving object, labeled $R$, located at $t,\chi$ is $a (\frac{\partial \chi}{\partial t})_R $.  \\

Definitions: We will use AP (almost physical) to describe spherically symmetric coordinate systems $x^{\mu}$ ($R,\theta,\phi,T$) that are transforms from the FLRW coordinates 
with a radial metric that approaches $\hat{M}$ as $\chi$ approaches zero.  We will attach the AP space origin to the same galactic point as $\chi = 0$,  so at this point there is no motion between them.  Thus we can call the AP coordinates  stationary.  
Since their differentials have a $\hat{M}$ metric close to the origin, they can be $\hat{L}$ transformed there from the physical coordinates $dt$ and $a(t)d\chi$.
For any point on $R$, they will have contravariant vectors  for velocity  $U ^{\mu} = \frac{dx^{\mu}}{ds}$ and acceleration $A^\mu = \frac{DU^\mu}{Ds}$, whose components transform like the coordinate differentials. $R$ is rigid in a mathematical sense because the radial component of $U^\mu$ in stationary coordinates is $(\frac{\partial R}{\partial T})_R(\frac{dT}{ds}) \equiv 0$, so the points of $R$ are motionless with respect to each other.  (It is rigid in its mathematical properties, but not in its acoustical properties).
It will be helpful in finding AP transforms if we further require the AP metric be diagonal (zero coefficient of $dTdR$).

We define a generalized Hubble ratio as $H(\hat{t}) = \frac{\dot{a}}{a}$, where the dot is the $d\hat{t} = c(t)dt$ derivative (eq \ref{3'}). \\

$iv$. ``The Lorentz transform between FLRW and AP radial coordinates is valid for the partial differentials of $T$ and $R$ from the origin out to at least the lowest power of the velocity between them.'' \\

Without this assumption, any light speed (includng constant) would be allowed \cite{F2}. \\

With these assumptions and definitions, we will show that the light speed is variable and proportional to 
$\dot{a} = aH$ (or equivalently to $\sqrt{\frac{da}{dt}}$) by two different procedures:\\

1) Integrate $\hat{L}$ transformed physical differentials $dT,dR$ in a power series in $\chi$. (Section \ref{pc}) \\

2) Find full rigid diagonal radial AP transforms $T,R$ for all $t,\chi$
(Sections \ref{cov}).\\

Each of these has the same variable light speed $c(t)$ in the limit of $\chi \rightarrow 0$. The first shows this for any and all AP transforms for an expansion of $T_t$ that is internally consistent to the second power of $\chi$ as required for Lorentz to be applicable.  The second shows this for a large number of full radial AP transforms which have a $\hat{M}$ metric close to the origin. Thus, the first is a completeness proof that if there are such transforms, they must have this $c(t)$, and the second is an existence proof that there are such transforms with $c(t)$ at the origin, and that the expansion of the first is further justified for being internally consistent to the second power of $\chi$.  \\

Additional assumptions are needed to apply this variable physical light speed to physical laws.  We will use the following:\\

($v$) ``We assume an extension of the Bernal criteria\cite{B} that one of two observers will have a physical coordinate when the other does if each calculates the other's differential of that coordinate at the same space-time point compared to his own and finds these cross measurements to be equal.''\\  

This seems reasonable since this is the characteristic of the transform between moving inertial frames.
We will find a radial AP transform $(T,R)$ called physical distance coordinates whose differential dR is physical for all distances by virtue of this assumption (Section \ref{cov} and Appendix \ref{pd}).  If the $dR$ represent readings on a stationary standard ruler, all on the same frame, they can be integrated to $R$ to determine a physical distance measured on a rigid physical ruler out into the far reaches of the universe.  \\

($vi$) ``Einstein's field equations can be maintained unchanged for $c(t)$ by assuming a gravitation ``constant'' that varies as $c(t)^4$.  This keeps constant the
 proportionality function between the GR field tensors (eq \ref{d9}). ''\\ 

The effect of $c(t)$ is introduced by an extended metric and an extended conserved stress-energy tensor (App \ref{gr}).  The extended FLRW metric solves the extended GR field equation for an ideal fluid.  A well-behaved transform will also be a solution since Riemann tensors are invariant to transforms.  The solution allows us to calculate $a(t)$ and $c(t)$ and galactic and photon paths on the AP frame for a homogeneous and isotropic universe with a variable light speed (Sects \ref{c}, \ref{paths}).\\ 

($vii$) ``We assume that the atomic frequency spectra of particles is constant.'' \\

Thus, we keep constant the fine structure constant and Rydberg frequency by making the vacuum electric and magnetic `constants' vary inversely with c(t).  This also allows us to redefine electro-magnetic field vectors to maintain Maxwells equations (App \ref{em}).\\

\subsection{Variable light speed c(t) required for a transform that is Lorentz close to the origin}\lbl{pc}

\subsubsection{Extended Lorentz transform from galactic points to the AP inertial frame using the velocity $V$ between them}\lbl{lor}

We will consider only radial world lines with physical coordinates $T$ and $R$ on the AP inertial frame.  We would like these to describe the same events as the FLRW coordinates $t$ and $\chi$ (App \ref{FLRW}), so $T = T(t,\chi)$ and $R = R(t,\chi)$ with $R = 0$ at $\chi =0$.
So
\begin{equation} \begin{array}{l}
dT = T_tdt + T_\chi d\chi = \frac{1}{c}T_td\hat{t} + T_\chi d\chi, \\[2mm]
dR = R_tdt + R_\chi  d\chi = \frac{1}{c}R_td\hat{t} + R_\chi  d\chi,
\end{array}  \lbl{7} \end{equation}
where the subscripts  
 indicate partial derivatives
with respect to  the subscript variable, and where we use $d\hat{t} = c(t)dt$ (Sect\ref{s2.1.2}).  
We will find $T = T(t,\chi)$ and $R = R(t,\chi)$ by integrating the differentials of the Lorentz transform for a short distance.   
We assume the $\hat{M}$ metric applies to physical differential times and distances of limited size anywhere and anytime.     
The FLRW metric in eq \ref{1} has a radial Minkowski-like metric with $dT^* \rightarrow dt$ and $dR^* \rightarrow ad\chi$ that we have assumed are physical.  
If a point on the AP frame is moving at a radial velocity $V(t,\chi)$ when measured with the FLRW coordinates, the $\hat{L}$ transform of $dt,adx$ to $dT,dR$ for a radial path keeps the line element $ds$ invariant (eq \ref{sr2}):
\begin{equation}
\begin{array}{c}
dT = \gamma (t,\chi) (dt - \frac {V(t,\chi)}{c(t)^2}a(t)d\chi) ,\\[2mm]
dR = \gamma (t,\chi) (-V(t,\chi)dt + a(t)d\chi), \end{array}
\lbl{m43'}
\end{equation}

If we compare eq \ref{m43'} with eq \ref{7}, we get 
\begin{equation}
T_t = \gamma ,  \lbl{m431} \end{equation}

\begin{equation}
T_\chi = -\gamma a\frac{V}{c}^2 = -\gamma a{\hat{V}}{c}, \lbl{m432}\end{equation}

\begin{equation}
R_t = -\gamma V = -\gamma c\hat{V} , \lbl{m433}\end{equation}

\begin{equation} R_\chi = \gamma a, \lbl{m434} \end{equation} 
where for simplification we have introduced $\hat{V} \equiv \frac{V}{c}$. These relations are exact for differentials as $\chi \rightarrow 0$, and therefor are approximately correct when the differentials are integrated for small $\chi$ at constant $t$.  We can rearrange the two expressions for $\hat{V}$ to give 
\begin{equation}
	\hat{V}  = -\frac{aR_t}{cR_\chi} = -\frac{cT_\chi}{aT_t}. \lbl{m45} \end{equation}

With eqs \ref{m431},  \ref{m434}, and \ref{m45} this gives two relations each for $dT$ and $dR$ in terms of $\hat{V}$.  When we integrate these partial differential equations, we integrate $dT,dR$ along the $R$ frame but integrate the $dt,d\chi$ along a radial connection between the co-moving galactic points $\chi$. Because the radial differential changes with time, $V(t,\chi)$ changes with time and distance.
We will find this combination requires $c(t)$ to vary with $t$ in a determined way, at least for the short distance from the origin where a power series is valid.
When there is no acceleration, and $V$ is a function only of $\chi$ in an expanding universe, $c(t)$ will be constant (see Appendix A.5).

\subsubsection{Power series in $\chi$ determines c(t)}\lbl{pc2}

To obtain $T(t,\chi)$ and $R(t,\chi)$ near the origin, we need to integrate the differentials $dT$ and $dR$ for small $\chi$. We will do this by expanding these physical coordinates in a power series in $\chi$ out to the lowest power that will give a non-trivial $c(t)$ in the limit of zero $\chi$.  
We will use the two relations for $dR$ to determine the expansion coefficients of $R$ and $\hat{V}$, then use the resultant expansion of $\hat{V}$ in the two relations for $dT$ to expand $T$ and determine the requirement for $c(t)$.  

Since $\hat{V}$ will vanish at the origin (see definitions, Sect \ref{ass}), the constant in the power series for $\hat{V}$ is zero; so let
\begin{equation}
	-\hat{V} = w_1(t)\chi + w_2(t)\chi^2 +  w_3(t) \chi^3+\bigcirc(\chi^4)...  
\lbl{p1}
\end{equation}
where the $w_i(t)$ are unknown functions to be determined.  From $R_\chi = a \gamma$ (eq \ref{m434}) we get
\begin{equation}
	R_\chi = a(1 + \frac {1}{2}\hat{V} ^2 + \frac{3}{8}\hat{V}^4 +...) = a(1 + \frac{1}{2} w_1^2 \chi^2 +w_1w_2\chi^3 + \bigcirc(\chi^4))...
\lbl{p3}
\end{equation}
If we integrate eq \ref{p3} at constant $t$, noting that $R$ vanishes at $\chi = 0$ (see definitions, Sect \ref{intro}), we obtain
\begin{equation}
	R = a\chi + \frac{1}{6}a w_1^2 \chi^3 +\frac{1}{4}aw_1w_2 \chi^4 + \bigcirc(\chi^5)...
\lbl{p4}
\end{equation}
$R(t,\chi)$ is the physical differential $dR$ summed over all the galactic points up to $\chi$, and is thus the physical distance to $\chi$ at time $t$.
The first term of eq \ref{p4} is the ``proper'' distance to which all measurements of distance reduce close to the origin \cite{We1}.

Partial differentiation of eq \ref{p4} by $t$ at constant $\chi$ gives
\begin{equation}
	R_t = c\dot{a}\chi + \frac{1}{6}\chi^3 \frac{d}{d t}(aw_1^2) + \frac{1}{4}\chi^4\frac{d}{d t}(aw_1w_2)  + \bigcirc(\chi^5)...
\lbl{p4'}
\end{equation}
where the dot represents the derivative with respect to $\hat{t}$.
We can then find $\hat{V}$ from eqs \ref{m45}, \ref{p3}, and \ref{p4'}:
\begin{equation}
	-\hat{V}  =  \frac{a R_t}{cR_\chi} = \dot{a}\chi + f(t)\chi^3+ \bigcirc(\chi^4)...,
\lbl{p5}
\end{equation}
where
\begin{equation}
	f(t) =-\frac{1}{2}w_1^2 \dot{a} +\frac{1}{6c} \frac{d}{d t} (aw_1^2)
\end{equation}
By comparison of eq \ref{p5} with eq \ref{p1}, we see that $w_1 = \dot{a}$, $w_2 = 0$, and $w_3 = f(t)$.  

We will now use this expression for $\hat{V}$ to find two relations for $T_t$.  The first comes from  $T_t = \gamma$ (eq \ref{m431}): 
\begin{equation}
	T_t = 1 + \frac {1}{2}\hat{V} ^2 = 1 + \frac{1}{2} \dot{a}^2 \chi^2 + \bigcirc(\chi^4)...,
\lbl{p2}
\end{equation}
Even though $dT$ and $dt$ are both measured on standard physical clocks, we note that the galactic clocks $t$ measured at constant $\chi$ run slower than the AP clocks $T$ as they move away from the origin (dilation).   
When measured at constant $R$, the AP clocks run slower, $t_T = 1+ \frac{V^2}{2}$ , in accordance with the Lorentz transform.  Neither of these apply if we don't carry out the power series to the second power of $\chi$.  Of course the distance contraction is also consistent with Lorentz, $\frac{R_\chi}{a} =a\chi_R = 1+ \frac{\hat{V}^2}{2}$ (eq \ref{p3}). 

We can find an expression for $T_\chi$, using eqs \ref{m45}, \ref{p2} and \ref{p5}:
\begin{equation}
T_\chi = -\frac{a}{c}T_t\hat{V} = \frac{a}{c}[1+\frac{1}{2}\dot{a}^2\chi^2 +\bigcirc(\chi^4)...][\dot{a} \chi + f(t)\chi^3 +\bigcirc(\chi^4)...],
\end{equation}
and multiplying the brackets gives
\begin{equation}
	T_\chi = \frac{a}{c}[\dot{a} \chi + \frac{1}{2}\dot{a}^3\chi^3 + f(t)\chi^3 +\bigcirc(\chi^4)...].
\end{equation}
By integration with $\chi$ at constant $t$ with $T = t$ at $\chi = 0$ we find
\begin{equation}
	T = t + \frac{1}{2}\frac {a \dot{a}}{c} \chi^2 + \bigcirc(\chi^4)... 
\lbl{p6}
\end{equation}

If we partially differentiate eq \ref{p6} by $t$, we get a second expression for $T_t$:
\begin{equation}
	T_t = 1 + \frac{1}{2}\chi^2 \frac{d}{dt}(\frac{a\dot{a}}{c}) + \bigcirc(\chi^4)...
\lbl{p7}
\end{equation}

If the transform is to have a Lorentz transform close to the origin, we must have the two expressions for
$T_t$ (eqs \ref{p2} and \ref{p7}) agree to at least the 2nd power of $\chi$ (i.e., the second power of $\hat{V}$).  This leads to a differential equation that determines a variable $c(t)$ given by
\begin{equation}
	 \dot{a}^2 = \frac{d}{dt}(\frac{a\dot{a}}{c(t)}).
\lbl{p8}
\end{equation}
Also mathematically, when we regard $c(t)$ as a variable to be determined by the limiting process of $\chi \rightarrow 0 $, we must keep the term in $\chi^2$ since it is the lowest term that determines $c(t)$, which we have therefor called non-trivial.  (Fletcher\cite{F2} shows that a transform with physical distance can be found for a constant light speed that leads to a $T_t \propto \frac{\hat{V}^2}{4}$, and therefor is not consistent with a Lorentz transform and is valid for only a smaller range of physicality).

To get an explicit expression for $c(t)$,  multiply eq \ref{p8} by $a$, change the variable $dt$ to 
$da = \dot{a} c(t)dt$ to yield
\begin{equation}
	\frac {da}{a} = \frac {c}{a\dot{a}} d(\frac{a\dot{a}}{c}).
\end{equation}
One can see that $c \propto \dot{a} $ is a solution, so
\begin{equation}
	\frac {c(t)}{c_0} = \frac{ \dot{a}(\hat{t})}{\dot{a}(t_0)} = \alpha E,
\lbl{p10}\end{equation}
where $\alpha$ is the normalized scale factor 
\begin{equation}
	\alpha \equiv \frac{a}{a_0}, 
\lbl{m40"}
\end{equation}
and $E$ is the normalized Hubble ratio $H(\hat{t})$
\begin{equation}
	E \equiv \frac{H}{H_0} = \frac{1}{H_0} \frac{\dot{a}}{a}.
\lbl{m40'}
\end{equation}
The subscript $0$ denotes the value at $t = t_0$, the present time.  We can take $c_0$ to be unity, so that $c(t)$ would be measured in units of $c_0$, but for most equations in this paper I will retain $c_0$ for clarity.  The field equation (sect \ref{c}) will enable us to evaluate  $\alpha$ and $E$  
and thus $c(t)$.

\subsection{Variable light speed c(t) derived from radial AP transforms (defined in sect \ref{ass})}\lbl{cov}

\subsubsection{Procedure for finding radial AP transforms using the velocity $V$}
We would now like to find radial AP transforms that will hold for all values of the FLRW coordinates and reduce to the physical coordinates for small distances from the origin. 
The most general line element for a time dependent spherically symmetric (i.e., isotropic) line element (Weinberg, p335\cite{We1}) is
\begin{equation}
	ds^2 = c^2 A^2 dT^2 - B^2 dR^2  - 2cCdTdR- F^2(d\theta^2 + sin^2 \theta d\phi^2)
\lbl{5'}
\end{equation}
where $A$, $B$, $C$, and $F$ are implicit function of $T$ and $R$, but explicit functions of $t$ and $\chi$.  We are using the same notation for time and distance as we did for the physical coordinates, but understand that they may be physical only for small distances from the origin.  We have included the physical light speed $c(t)$ in the definition of the coefficient of $dT$.

We will look for transformed coordinates which have their origins on the same galactic point as $\chi =0$, so $R = 0$ when $\chi = 0$, where there will be no motion between them, $\hat{V} = 0$, and where $T $ is $t$, since the time on clocks attached to every galactic point
is $t$, including the origin.  We will use the same angular coordinates as FLRW and make $F = ar$ to correspond to the FLRW metric, but will find only radial transforms where the angular differentials are zero.  Of course, full four dimensional transforms to time and three rigid axes have not been found, nor are they required to determine $c(t)$.  They have only to meet the requirement of becoming $\hat{L}$ close to the origin.  By definition radial AP transforms do this.

Then $R$ and $T$ will be functions of only $t$ and $\chi$: $T = T(t,\chi)$ and $R = R(t,\chi)$, and we will still have eq \ref{7}.
Let us consider a radial point at $R$ in the AP system.  When measured from the FLRW system ($\chi,0,0,\hat{t}$), it will be moving at a velocity given by
\begin{equation}
 V = a(t)\left( \frac{\partial \chi}{\partial t} \right)_{\! R} = c\hat{V}, 
\lbl{12}
\end{equation}
This velocity will be the key variable that will enable us to obtain radial AP transforms of the full radial coordinates.  

We will now find the components of the contravariant velocity vector $U^{\hat{t}} = \frac{d\hat{t}}{ds}$ of a point on the $R$ axis in both the FLRW coordinates and the AP coordinates.
To get the time component in FLRW coordinates $\chi, \theta, \phi,\hat{t}$ we divide the FLRW metric (eq \ref{3"}) by $d\hat{t} ^2$ with $d\omega = 0$ to obtain
\begin{equation}
	(\frac {ds}{d\hat{t}})^2 = 1 - a(t)^2 (\frac{d\chi}{d\hat{t}})^2 = (1 - \hat{V}^2) \equiv \frac{1}{\gamma^2}
\lbl{12'}
\end{equation}

To get the spatial component, we use the chain rule applied to eqs \ref{12} and \ref{12'}:
\begin{equation}
	\frac{d\chi}{ds} = \frac{d\chi}{d\hat{t}} \frac{d\hat{t}}{ds} = \frac{\hat{V}}{a}\gamma.
\end{equation}

For AP coordinates $R, \theta, \phi, T$, the radial component of the contravariant velocity vector is zero (see definitions under assumption iii).  The point is not moving in those coordinates; that is, the radial component is rigid.  This means that a test particle attached to the radial coordinate will feel a force caused by the gravitational field, but will be constrained not to move relative to the coordinate.  Alternatively, a co-located free particle at rest relative to the radial point will be accelerated, but will thereafter not stay co-located.  

The AP time component of the velocity vector is $\frac{dT}{ds} = \frac{1}{cA}$.  This makes the vector 
$U^{\mu} = \frac {dx^{\mu}}{ds}$ in the AP coordinates 
\begin{equation}
	U^{\mu}= (0,0,0,\frac{1}{cA})
\lbl{u1}\end{equation}
 and in the FLRW coordinates 
\begin{equation}
	U^{\mu} = (\frac{\gamma \hat{V}}{a},0,0,\gamma).
\lbl{u2}\end{equation}  To make it contravariant, its components must transform the same as $dT,dR$ in eq \ref{7}:

\begin{equation}
\begin{array}{rcl}
\frac{1}{cA}& = & \frac{1}{c}T_t\gamma + \frac{1}{a}T_\chi \gamma \hat{V},\\[.15cm]
0 & = & \frac{1}{c}R_t \gamma + \frac{1}{a} R_\chi \gamma \hat{V}. \end{array} \lbl{11}.
\end{equation}

Manipulating the second line of eq \ref{11} gives
\begin{equation}
\hat{V}  =  -\frac{a R_t}{cR_\chi}. \lbl{m13}\end{equation}

If we invert eq \ref{7}, we get
\begin{equation}\begin{array}{l}
d\hat{t} = \frac{1}{D}(R_\chi dT - T_\chi dR),\\[2mm]
d\chi = \frac{1}{D}(- \frac{1}{c}R_t dT + \frac{1}{c}T_t  dR),
\end{array}  \lbl{iv}
\end{equation}

where 
\begin{equation}
 	D = \frac{1}{c}T_tR_\chi - \frac{1}{c}R_tT_\chi = \frac{1}{c}T_tR_\chi (1 + \hat{V}\frac{cT_\chi}{aT_t}),
\lbl{iv2}
\end{equation}
using eq \ref{m13}.  
We can enter $d\hat{t}$ and $d\chi$ of eq \ref{iv} into the FLRW metric (eq \ref{1})to obtain coefficients of $dT^2$, $dR^2$ and $dTdR$.  One way to make $ds^2$ invariant is to equate these coefficients to those of eq \ref{5'}:
\begin{equation}
	A^2 = \frac{1}{T_t^2} \left[ \frac {1-\hat{V}^2}{(1+\hat{V}\frac{cT_\chi}{aT_t})^2}\right],\lbl{in1}
\end{equation}
\begin{equation}
	B^2 = \frac{a^2}{R_\chi ^2} \left[ \frac{1-(\frac{cT_\chi}{aT_t})^2}{(1+\hat{V}\frac{cT_\chi}{aT_t})^2}\right],\lbl{in2}
\end{equation}
and
\begin{equation}
	C = -\frac{a}{T_tR_\chi } \left[ \frac{\hat{V}+\frac{cT_\chi}{aT_t}}{(1+\hat{V}\frac{cT_\chi}{aT_t})^2} \right].\lbl{in3}
\end{equation}
If we put $ds=0$ in eq \ref{5'}, we obtain a coordinate velocity of light $v_p$:
\begin{equation}
	\frac {v_p}{c} =  \left[\frac{\partial R}{\partial\hat{T}}\right]_s = -\frac{C}{B^2} \pm \sqrt{(\frac{C}{B^2})^2 +  \frac{A^2}{B^2}}
\lbl{2'}
\end{equation}
We need to remember that the $c(t)$ in these equations is the physical light speed assumed for the FLRW metric.  

The equations for $A$, $B$, and $v_p$ simplify for a diagonal metric ($C = 0$).  Then eq \ref{in3} becomes
\begin{equation} 
\frac{cT_\chi}{aT_t} = -\hat{V}
\lbl{diag}\end{equation}
and eqs \ref{in1}, \ref{in2}, and \ref{2'} become
\begin{equation}
	A = \frac{\gamma}{T_t} = \frac{t_T}{\gamma}
\lbl{in1'}\end{equation}
\begin{equation}
	B =  \frac{a\gamma}{R_\chi} = \frac{a\chi_R}{\gamma}
\lbl{in2'}\end{equation}
\begin{equation}
	\frac {v_p}{c} = \frac{A}{B},
\lbl{in2"}\end{equation}
where we have used eq \ref{iv} with $C = 0$ to obtain the inverse partials.  

Thus, rigidity gives us a relation of $dR$ to $\hat{V}$ (eq \ref{m13}), and diagonalization gives us a relation of $dT$ to $\hat{V}$ (eq \ref{diag}).  If we find $\hat{V}(t,\chi)$, we can find $R(t,\chi)$ and $T(t.\chi)$ by partial integration.

This metric becomes $\hat{M}$ when $A \rightarrow 1, B \rightarrow 1, C \rightarrow 0 $ and $ar \rightarrow R$, and we get the relations in eqs \ref{m431}-\ref{m434} so that the transformed metric  becomes $\hat{M}$ in four dimensions.  The light speed for AP coordinates differs from that of the FLRW coordinates as $R$ increases from zero by the ratio $\frac{A}{B}$.

Even when the  full physicality conditions are not met, we can say something about the physicality of the coordinates with a generalization of criteria ($iv$) developed by Bernal et al\cite{B}.
They developed a theory of fundamental units based on the postulate that two observers will be using the same units of measure when each measures the other's differential units at the same space-time point compared to their own and finds these reciprocal units to be equal.  We generalize this by stating that if one coordinate of a system represents the reading on a physical instrument, so must the corresponding coordinate of the other  reciprocal system represent readings on the same type of physical instrument with the same units.  
Thus, 
if $A,B,C = A,1,0$, $dR$ will be physical  because  $\frac{R_\chi}{a} = a \chi_R = \gamma$ (eq \ref{in2'}) and $dR$ uses the same measure of distance as $ad\chi$, which FLRW assumes is physical.  
If the $dR$ represent readings on a stationary standard ruler, all on the same frame, they can be integrated to $R$ to form a rigid physical ruler out into the far reaches of the universe.  
(Of course, the converse is not true; if this equality does not hold for $dT$, it may still be physical, but the clocks may be running slower due to gravitational time shifts; e.g., see eq \ref{vB3}).  Similarly, if $A,B,C = 1,B,0$, then the AP transform will have physical time.

At this point we would like to examine quantitatively how far from the $\hat{M}$ metric our transformed metric is allowed to be in order for its coordinates to reasonably represent physical measurements.  We can consider the coefficients $A$, $B$, and $C$ one at a time departing from their value in the $\hat{M}$ metric. For example, let us consider the physical distance case $B=1,C=0$ and examine the possible departure of the time rate in the transform from that physically measured. Then, from eqs \ref{in1'}: $T_t = \frac{\gamma}{A}, t_T = \gamma A$.  
Thus, $1-A$ represents a fractional increase from $\gamma$ in the transformed time rate $T_t$ and $dT$, and thus the fractional increase from physical of an inertial rod at that point.  
We can make a contour of constant $A$ on our world map to give a limit for a desired physicality of the transform.\\\

\subsubsection{Diagonal radial AP transformed coordinates have physical c(t) close to the origin} \lbl{gs}

We show in Appendix \ref{ct} that there exist an infinite number of radial AP transformed coordinate systems which satisfy the $\hat{M}$  requirements close to the origin.  Appendix \ref{pt} derives diagonal transforms ($C = 0$) using physical time ($A = 1$) for all physical times $T$. These all independently show that the light speed becomes $c(t) \propto \dot{a}$ for small distances, where the transforms become Lorentz. 

Appendix \ref{pd} shows the diagonal transforms for physical distance ($B = 1, C = 0$) for all physical distances $R$.  To integrate the PDEs for this transform, we need to use the GR Field Equation (FE).  Because the equations in \ref{pt} and \ref{pd} are different from each other, they show, as we would expect, that it is not possible to have diagonal transforms with  physical $R$ and physical $T$ simultaneously for all values of $t,\chi$ (except for an empty universe).  

At all distances for $A=1,C = 0$, the AP time $T$ can be measured on AP physical clocks, but the AP distance $R$ cannot be measured on physical rulers for all distances.  
For $B=1, C=0$, the AP distance $R$ can be measured by physical rulers on a AP frame for all distances, but the AP time $T$ cannot be measured by physical clocks (except for small $R$).  
We can calculate an acceleration (Appendix \ref{s2.1.1}) for a flat universe that is zero at the origin, and increases with distance; the physical distance $R$ acts like you might expect for a rigid ruler on whom the surrounding masses balance their gravitational force to zero at the origin, but develop an inward pull as the distance increases.

Appendix \ref{fu} describes similarity solutions for both types for a flat universe ($\Omega = 1$).  These solutions are very useful to display alternatives.  For the Physical Distance transform, when we use the FE for a constant light speed \cite{F2}, we get a transform that does not have the Lorentz dependence on $\hat{V}$.  When we use the FE that allows a varying light speed, this yields a transform that has the Lorentz dependence on $\hat{V}$ if, and only if, we use the same light speed $c(t)$ as for the power series expansion and the Physical Time transform.  This self-consistency indicates that we are using the correct FE and the correct $c(t)$.  

To summarize, we have shown that every transform that has a variation of $T_t = 1 + \frac{\hat{V}^2}{2}$, as required by a Lorentz transform close to the origin, has $c(t) \propto \sqrt{\frac{da}{dt}}$.  If we do not require this variation of $T_t$, it is possible to find a physical distance transform with a constant $c$ \cite{F2}, although its physicality goes a much shorter distance into the universe.  However, it is not possible to find a diagonal physical time transform with constant $c$ (see App \ref{pt}).  Although there is no requirement that there be such a transform nor that the physical distance transform have a large range of physicality, the derived $c(t)$ has an attractive universality that can be made consistent with special and general relativity (see sect \ref{c} and App \ref{s2.1.2}).

\section{Extension of GR to incorporate c(t)} \lbl{c}
We can accommodate the variable light speed $c(t)$ in the field equation of general relativity for FLRW by allowing the gravitational ``constant'' $G$ to be time varying so as to keep constant 
the proportionality function between the GR tensors (eq \ref{d9}).  We avoid taking derivatives of $c(t)$ by using $\hat{t}$, where $d\hat{t} = c(t)dt$.
The dependence on real time $t$ is found by transforming the resultant solution back to $t$ from $\hat{t}$.  This is described in App \ref{FE}.  This enables 
us to calculate $a(t)$, $c(t)$, and trajectories in the time and distance of AP coordinates.

In eqs \ref{d6},\ref{d7} of the Appendix are the two significant field equations of
 the extended GR applied to an ideal fluid:
\begin{equation}
	\frac{3 \dot{a}^2}{a^2} + \frac{3k}{a^2} - \Lambda = \frac{8 \pi G}{c^4} \rho c^2,
\lbl{f6}
\end{equation}
and
\begin{equation}
	+2 \frac{\ddot{a}}{a} +\frac{\dot{a}^2}{a^2} +\frac{k}{a^2} -\Lambda = -\frac{8 \pi G}{c^4}p,
\lbl{f7}
\end{equation}
where the dots represent derivatives with respect to $\hat{t}$.  
Following Peebles \cite[p312]{P}, we define
\begin{equation}
	\Omega \equiv \rho _0  \frac{8\pi G_0 }{3c_0^2H_0^2}  
\lbl{a3}
\end{equation}
and
\begin{equation}
	\Omega _r \equiv \frac{-k}{H_0 ^2a_{0}^2}
\lbl{a4}
\end{equation}
and
\begin{equation}
	\Omega _{\Lambda} \equiv \frac {\Lambda}{3 H_0 ^2}.
\lbl{a5}
\end{equation}
For very small $a$ there will also be radiation energy density which will not be considered in this paper.
 
The normalized Hubble ratio $E$ in eq \ref{m40'} is determined by eq \ref{f6}:
\begin{equation}
	\frac{1}{H_0}\frac{\dot{a}}{a} = E = \sqrt{\frac{\Omega}{\alpha ^3} + \frac {\Omega _r}{\alpha ^2} + \Omega _{\Lambda}}.
\lbl{a7}
\end{equation}
which allows us to evaluate $\frac{c(t)}{c_0} = \alpha E$.
The $ \Omega $s are defined so that
\begin{equation}
	 \Omega + \Omega_r + \Omega _{\Lambda} = 1.
\end{equation} 
At $t = t_0$: $\alpha = 1$, $E = 1$, and $\frac{c}{c_0} = 1$.

The cosmic time $t$ measured from the beginning of the FLRW universe (Big Bang) becomes
\begin{equation}
	c_0 H_0 t = \int_0^{\alpha}\frac {c_0 d\alpha}{c\alpha E} = \int_0^{\alpha} \frac{d\alpha}{\alpha ^2 E^2}.
\lbl{a9}
\end{equation}

For a flat universe with $\Omega = 1$ and $\Omega_r = \Omega_{\Lambda} = 0$: 
\begin{equation} \begin{array}{l}
c_0 H_0 t = \frac{\alpha ^2}{2}, \\[1.5mm] 
c_0 H_0t_0 = \frac{1}{2}, \\[1.5mm]
\alpha = (\frac{t}{t_0})^{\frac{1}{2}}\end{array} \end{equation}
\begin{equation}
	\frac{c}{c_0} = \alpha E = \alpha ^{-\frac{1}{2}} = (\frac{t_0}{t})^{\frac{1}{4}}.\end{equation}
For other densities with $\Omega_r = 1-\Omega $, $\Omega_{\Lambda} = 0$, 
\begin{equation}
c_0 H_0 t = \frac{\Omega}{(1-\Omega )^2}[y-\ln{(1+y)}], \lbl{a11}\end{equation}
 where
\begin{equation}	y = \frac{1-\Omega }{\Omega}\alpha. \lbl{a12}\end{equation}	
There is no periodicity of $\alpha$ with $t$ for $\Omega > 1$, .  The higher density decreases the time 
$t_0 \rightarrow \ln{\frac{\Omega}{c_0 H_0}} \Omega$, and the universe scale factor $\alpha$ continues to expand, asymptotically approaching a maximum at
$\frac{\Omega}{ (\Omega - 1)}$.  As $\Omega \rightarrow 0, a \rightarrow t, c \rightarrow c_0  $, the universe becomes Minkowski (see Appendix A.5).

For experiments attempting to measure the variation of the light speed at the present time, the derivative of $c(t)$ (eq \ref{p10} with $\frac{\Omega _b}{\alpha ^4 }<< 1$) will be more useful:
\begin{equation}
	\frac{1}{c_0H_0}\left[ \frac{1}{c} \frac{dc}{dt}\right] _{t=t_0} = 1- \frac{3}{2}\Omega -\Omega _r = -\frac{\Omega}{2} + \Omega _{\Lambda}.
\lbl{a10}
\end{equation}
Notice that this fraction is negative when matter dominates, and goes from zero at zero density to $-\frac{1}{2}$ at the critical universe density.  A vacuum energy density opposes the gravitational effect of matter; when it dominates, the slope is an increasing function of time. \\\

\section{Paths of galactic points and received light} \lbl{paths}

Because there is a special interest in having a physical description for distance in the universe, we display the physical distance transforms.  The physical distance results for flat space (Appendix \ref{fu}) are shown  in Figs. 1 and 2.  Here we have used the field equations with the generalized time (Sect \ref{c}) to derive the equations for $a(t) = a_0(\frac{t}{t_0})^{\frac{1}{2}}$ and $c(t) = c_0(\frac{t_0}{t})^{\frac{1}{4}}$.  

Fig 1 plots distance $R$ against the time at the origin (cosmic time $t$) for galaxies (constant $\chi$) 
and for incoming light reaching the origin at $\frac{t}{t_0} = 1$.  The galactic paths are labeled with their red shift $z$, determined by the time $t$ of the intersection of the photon path with the galactic path $z = -1 +c/\alpha = -1 + (t_0/t)^{3/4}$, assuming the frequency of the emitted light does not change with $c(t)$.   Notice that light comes monotonically towards the origin from all galactic points.  This photon path has a slope of $c_0 = 1$ close to the origin where the distance $R$ and time $t$ are both physical, but decreases as the distance increases and the time decreases, different from $c(t)$.  

Although the distance uses physical rulers, the coordinate system as a whole may not be physical for times shorter than some limit. A reasonable limit (sect \ref{gs}) might be $A = 0.95$, 
$R = 2.3(t/t_0)^{3/4}$, shown by the heavy dotted line in the figures.  
Together with $B=1$,$C=0$ for these physical distance coordinates, this shows that the assumption that $T$ and $R$ represent a physical AP coordinate system inside this limit is very good, with distances accurately represented, and
time rates $T_t$ within $5\%$ of physical measurements on adjacent inertial rods.  

Fig 2 plots these distances vs the transform time $T$ at $R$. At the emission of the photons, $T$ is finite (even for $t = 0$), presumably the transformed time it takes for the galactic point to get out to the point of emission.  
 At $\frac{T}{t_0} = 1$ the slope of the light path is $c_0 = 1$, and at the physicality limit $\frac{T}{t_0}=0.40$ the slope is only $5\%$ less than $c(t) = 1.50$.
  At the intersection of this physicality limit with the photon path that arrives at the origin at $t_0$, the time $\frac{t}{t_0} = 0.2 $ and the red shift $z = 2.4 $.  Thus, if we have a flat universe with $\Omega = 1$, the last $80\%$ of the universe history out to a $z$ of $2.4$ can be treated with  physical coordinates $T$ and $R$.  This $z$ is as large as any of the Supernova Ia whose measurements have suggested an accelerating universe.  
It extends out into the universe much farther than a similar transform for a constant light speed\cite{F2} that extends only out to a red shift of $z = 0.5$.

When the velocity of the points of the physical distance approaches the light speed when viewed from FLRW, the physical distance shows a Fitzgerald-like contraction so that it reaches a finite limit at the horizon ($t = 0$), beyond which there are no galactic points and no space.  This is true for all universe densities including an empty universe.  (It is also true for a constant light speed\cite{F2}).

I have included three additional figures, also using the extended field equation of section  \ref{c}  (and sect \ref{FE}).  Fig 3 is for a density of $\Omega = \frac{1}{2}$ (Appendix \ref{pd}), which has paths intermediate between $\Omega = 1$ and $\Omega = 0$. Fig 4 shows the effect of dark energy (Appendix \ref{pd} for $\Omega _{\Lambda} = \frac{3}{4}$), where all the curves tend to have inflection points when the dark energy becomes dominant.  The empty universe ($\Omega = 0$ in Appendix \ref{pt.5}) shown in Fig 5 is physical for all space-time, undistorted by gravitational curvature; galactic points and light travel in straight lines.  It is very similar to Figs 2-4 in that it demonstrates a finite horizon, beyond which there are no galactic points and no space. Figs 1-2 are from the numerically integrated similarity solution, Figs 3-4 are from the numerically integrated initial value solution, and Fig 5 is an analytic function solution\cite{R2}.  These illustrate complete coverage of $0 \leq \Omega \leq  1$.

\section{Underlying physics}\lbl{phys}

Our objective of transforming the FLRW into the physical variables of the AP frame is the same as Zelmanov's chronometric invariants\cite{Z} that project events onto observable coordinates.  The AP transforms, of course, do not have the generality of Zelmanov's chronometric invariants.  Somehow, the variable light speed $c(t)$ considered as physical according to my definition must be a variable function of his invariant constant 
observable light speed.  The present paper allows for the possibility of a variable light speed and then derives a relation for it for the FLRW universe.  The GR field equation can be maintained unchanged to calculate value for $c(t)$ as a function of the universe energy density and curvature by assuming the gravitational constant $G$ to be proportional to $c(t)^4$.

It really should not surprise us that the universe has a variable light speed.  It is well known that an observer accelerated relative to an inertial observer measures a 
 variable light speed depending on the acceleration.
(see MTW\cite{MTW}p173).  

The effect of gravitational potential on light speed is also demonstrated by
the Schwarzschild coordinates, where the coordinate light speed as well as the time on clocks are changed by the gravitational potential at a distance from a central mass.   
 
In the FLRW universe there are clearly gravitational forces caused by the energy density of the universe.  These cause the expansion of the universe to be slowed down (or speeded up if dark energy predominates) shown by the change in the FLRW scale factor $\dot{a}(t)$.  
The case we have considered differs from either of the first two.  We have examined a rigid radial rod whose gravitational force as felt by an observer attached to the rod increases with distance along the rod.  The light speed $v_L$ measured by such an observer stays within 5\% of $c(t)$ while the latter changes by a factor of 1.5 (for $\Omega = 1$) out to the physicality limit. The variation is not directly caused by the acceleration $\frac{dV}{dt}$, but mostly by the change in $a$, which in turn is affected by the gravitational forces.  An alternate way to view the light speed variation is to recognize that the FLRW metric has already accounted for both the gravitational forces and the light speed variation when it satisfies the GR field equation, which therefor relates the two.

In Appendix C for FLRW we show that for a flat universe ($\Omega = 1$) with the presently derived variable light speed, there is a gravitational field $g$ in the physicality region that increases linearly with distance from the origin. 
If we insert into eq \ref{gB1} the mass of the universe inside the radius $R$, $M_0 = 4\pi \rho_0 \frac{R^3}{3}$ at time $t_o$, we obtain
\begin{equation}
	g = -\frac{G_0M_0}{R^2}
\end{equation}
the Newtonian expression for gravitational field at a radius $R$ inside a sphere of uniform density. This is another indication that the T,R coordinates are obeying special relativity laws near the origin because an accelerated particle in the rest frame of SR has the Newtonian acceleration\cite{MSL}.  Note that $g < 0$ indicates an inward pull on the galactic points towards the origin of the AP axis, which we can interpret as the cause for the universe expansion to slow down (for $\Lambda = 0$).  

Thus, just as the assumption of homogeneity requires the universe to be either expanding or contracting, it seems to require the physical light speed to depend on this rate of expansion or contraction.

\section {To observe c(t)} \lbl{ob}

The most straight forward way to observe $c(t)$ would appear to be a direct measurement of the light speed or of an atomic spectra wavelength with the same precision and stability that we can now measure spectra frequency.  A fractional change in speed or wavelength should be $6 x 10^{-17}$ in 100 secs or $2x10^{-11}$ in a year if $\frac{c}{c_0} = (\frac{t}{t_0})^{\frac{1}{4}}$.  With this much sensitivity, however, an observation would have to separate out the possible effects on light speed of the gravitational forces of local masses like the earth, the moon, and the sun.  

We also need a ruler whose dimensions do not change with time.  
 Thus, we can't use a ruler proportional to the wavelength of a standard spectrum line because it will be proportional to $c(t)$.  Even the standard platinum meter stick can change with $c(t)$ because the Bohr radius will be proportional to $c(t)$ (if our assumptions about $E/M$ are correct). 

Effects of $c(t)$ should be seen on observation of distant astronomical events.  For instance it will affect the use of supernova Ia\cite{P'}\cite{R1}\cite{RInv}\cite{T}  to measure the acceleration of distant galaxies. 
Other astronomical observations that might be affected by $c(t)$ are cosmic background radiation, gravitational lensing, and dynamical estimates of galactic cluster masses.  

Unfortunately, the $c(t)$ calculated herein solves neither the flatness nor the horizon problem without inflation:  The flatness problem changes little because the Hubble ratio has a similar dependence on the universe scale factor $a(t)$.  The horizon problem remains because $c(t)$ enters both the transverse speed of light and the radial speed of galactic points. At the time of the release of the CBR photons, without inflation light could have traveled laterally only $\theta = \int_0^{t}\frac{c\partial t_\chi }{r a} $, where $\chi = \int_t^{t_0} \frac{c dt}{a} $.                                    
For $\Omega = 1, z=3000$, $ \theta = \frac{1}{z^{1/3}} =.07$ radians, or $4$ degrees,  and so galactic points could not have interacted separated by more than this angle.  \\

\section{Conclusions} \lbl{conc}

From the cosmological principle of spatial homogeneity and isotropy we can obtain the FLRW metric, which allows a variable light speed, that 
describes a universe of inertial frames attached to expanding galactic points with FLRW differential co-moving coordinate times the scale factor $a(t)$ interpreted as a physical differential distance.  The FLRW metric is Minkowski-like in its radial derivative.  Locally, 
SR applies, so a AP rigid frame attached to the origin has a Minkowski metric.  Thus, for a radial world line we can use a Lorentz transform from FLRW to the AP frame that keeps the two Minkowski world line elements invariant in order to obtain time and distance coordinates to describe radial movement in the universe close to the origin.  Because the FLRW metric has a time varying coefficient multiplying the space differential, this produces a velocity between a galactic points and the AP frame that is a function of time and distance.  If the Lorentz transform is to remain valid out from the origin to the lowest power of this velocity, 
a consistent limiting process to zero distance from the origin 
requires a variable light speed $c(t) \propto \sqrt{\frac{da}{dt}}$), the square root of the rate of change of the scale factor of the FLRW universe.  

By homogeneity, the origin can be placed on any galactic point, so that this variable light speed enters physical laws throughout the universe.  

 We extend the field equation by allowing the gravitation ``constant'' and the rest masses of particles to vary in such a way as to keep constant the rest mass energy and the Newtonian gravitational energy.  We have shown that this results in a constant relating the tensors of the field equation, like the field equation with a non-variable light speed. 
This yields a new function of cosmological time for the scale factor of the FLRW universe and thus values for $c(t)$.  These enable the calculation of physical distance vs physical time for galactic and light paths 
in the universe.

Although three orthogonal rigid axes are inadequate to describe three dimensional motion in accelerating fields, it is possible to describe one dimensional motion on a single axis.  We have done this for the FLRW universe by finding radial AP transforms from FLRW for all distances whose differentials remain close to SR Minkowski with this same variable light speed out to a red shift of  $ 2$ for a flat universe.

I have shown that the physical coordinates on the AP frame near the origin have a gravitational field for a flat universe that increases linearly with radius just like the Newtonian field for a spherical distribution of uniform mass density.  
 Like Schwarzschild, a gravitational red shift is predicted for a distant AP light source observed at the origin of the FLRW universe.  

To summarize, I am persuaded that the physical light speed throughout the FLRW universe is proportional to $\sqrt{\frac{da}{dt}}$ because (1) based on usual assumptions, for small distance from the origin a radial Lorentz transform from FLRW to a AP rigid frame
requires it; (2) all radial AP transforms from FLRW coordinates that I have investigated that have a Lorentz transform from FLRW near the origin have this same variable light speed; (3) we can use the standard GR field equation (with $G \propto c^4 $) to calculate the transformed distance vs time for galactic points and light that behave in a physically sensible way; (4) the transformed gravitational field in the physicality region for a flat universe is Newtonian for a spherical distribution of uniform mass density and can be considered the cause of the deceleration of the universe 
(when dark energy can be neglected)  (5) the AP transform extends physicality out into space much farther than for a comparable transform with constant light speed.

Just as the assumption of homogeneity requires the universe to be either expanding or contracting, it seems to require the light speed to depend on this rate of expansion or contraction under the influence of gravity. 

One of the radial AP transforms from FLRW has a distance coordinate that remains physical for all distances.  We can interpret this to be a global reference distance (used in Figs 1-5), although the time of this transform becomes unphysical at large distances.

Some other physical ``constants'' that depend on the light speed must also be changing with cosmic time.  I have suggested some constraints on this variability: (1) retaining the conservation of the stress energy tensor, including keeping constant the rest mass energy, the gravitational energy, and the Schwarzschild radius, and (2) keeping frequency of atomic spectra constant, which means the fine structure constant, and the Rydberg frequency.  These still make possible the geometrization of relativity with an adaptation of vectors and tensors such as the energy-momentum vector, the stress-energy tensor, and the electromagnetic field tensor.  

A variable light speed should be observable by direct measurement of light speed or spectral wavelength with  clocks and rulers whose units remain constant, if they could be measured to the same precision as frequency, and if the possible effects on light speed of the gravitational forces of nearby  masses like the earth, the moon, and the sun could be isolated.  It should have an impact on understanding distant cosmic observations; e.g., analysis of the apparent acceleration of galaxies via supernova Ia, cosmic background radiation, gravitational lensing, and dynamical estimates of galactic cluster masses could all be affected.  
But the recognition of this $c(t)$ does not solve the flatness nor horizon problems without inflation.

I have outlined in Appendix \ref{s2.1.2}  how a variable light speed can be included in an extended special and general relativity by keeping constant the rest energy of particles and the energy of Newtonian gravity acting between them.

\section*{Appendix}

\appendix

\addcontentsline{toc}{section}{Appendix}

\section{AP (almost physical) coordinates with diagonal metrics} \lbl{ct}

\subsection{Physical Time}\lbl{pt}

\subsubsection{ Partial differential equation for $\hat{V} = \frac{V}{c(t)}$}

We will be considering radial AP transforms for diagonal coordinates that eq \ref{in3} makes 
\begin{equation}
	\hat{V}  = -\frac{cT_\chi}{aT_t}. \lbl{dc}\end{equation}

For diagonal coordinates with physical time at all $t$ and $\chi$, $A=1$. Thus, eq \ref{in1} becomes
\begin{equation}
	T_t = \gamma.
\lbl{m25}
\end{equation}
This automatically guarantees the Lorentz time dilation $(\frac{\partial T}{\partial t})_R = \frac{1}{t_T} = \frac{1}{\gamma}$ (eq \ref{iv}).  We need only find a transform for which $B \rightarrow 1$ close to the origin to make it AP.

We proceed by finding a differential equation with $\hat{V}$ as the only dependent variable.  
Thus, we can write 
a formula for $T$, using eq \ref{m25} and eq \ref{dc}:
\begin{equation} T = t + \int_0^\chi T_\chi\partial\chi_t = t + \int_0^\chi (-\frac{a}{c} \gamma \hat{V})\partial\chi_t, 
\lbl{20} 
\end{equation}
where we have used the boundary condition that at $\chi = 0$, $T = t$, and the symbol $\partial x_t$ signifies integration with $\chi$ at constant $t$.  It can be partially differentiated with
respect to $t$ (giving $\gamma $) and then with respect to $\chi$ and with the use of eq \ref{12}, noting that $d\gamma = \gamma^3 \hat{V} d\hat{V}$ and $1+\hat{V}^2\gamma^2 = \gamma^2$, we obtain
a PDE for $\hat{V}$:
\begin{equation}
	\hat{V}_t + \hat{V}_\chi \left(\frac{\partial\chi}{\partial t}\right) _R  = \left( \frac{\partial \hat{V}}{\partial t}\right)_{\! R}  = -\hat{V}(1 - \hat{V}^2)\frac{c}{a} \frac {d}{dt} (\frac {a}{c}). 
\lbl{23}
\end{equation}

\subsubsection{The general solution for $\hat{V}$, $R$. and $T$ for all $a(t)$} 

Eq \ref{23} can be rewritten as
\begin{equation}
	\frac{\partial \hat{V}_R}{\hat{V}(1 - \hat{V}^2)} = -\frac{\partial(\frac{a}{c})_R}{\frac{a}{c}}, \lbl{m30}\end{equation}
where the subscript on the partial differential indicates the variable to be held constant.  This can be integrated with an integration constant $\ln{\kappa}$. Since the integration is done at constant $R$, then 
$\kappa = \kappa(R)$, and inversely,
$R = R(\kappa)$.  
Integrating eq \ref{m30}, we get
\begin{equation}
\hat{V} = -\frac{\kappa}{\sqrt{\frac{a^2}{c^2} + \kappa^2}},\lbl{m31'}\end{equation}
where the sign of $\kappa$ will be positive for an expanding universe, where the $\chi$ points will stream out radially past a point at $R$. 

At this
point, $R$ is an unknown function of  $\kappa$. The various
possible 
coordinate systems which solve our PDEs are 
characterized, in large part, by the function $R(\kappa)$.
 But for all, in order for $\hat{V}$ to  vanish when
 $R = 0$ (see definitions sect \ref{intro}), $\kappa$ must also;  so always
\begin{equation}
\kappa(0)  = 0. \lbl{q60}\end{equation}
We note that as long as $\kappa(R)$ remains finite, $\hat{V}$ goes to $-1$, and $V$ goes to $-c(t)$, for
$a(t) = 0$, i.e. 
for $t = 0$, the horizon.   

Let us now look at lines of constant $\kappa(R)$, i.e.\
constant $R$, in $t,\chi$ space. 
Eq \ref{12} can be integrated for $\chi$
 with use of eq \ref{m31'} at
constant $\kappa$ to give the following:
\begin{equation}
\chi(t,\kappa) =  \int_t^{\infty}\frac{c\kappa \partial s_\kappa}{a(s)\sqrt{\frac{a^2}{c^2} +
\kappa^2}}. \lbl{m33}\end{equation}
 For an expanding universe, we have set the upper limit
 at $\infty $, because we expect that if $R$ is kept constant the galactic point $\chi$ that will be passing any given $R$ will eventually approach zero as FLRW time $t$ approaches infinity.  

At this point, we have obtained $\hat{V} = \hat{V}^{*}(t,\kappa)$
from eq \ref{m31'} and have also obtained the function
$\chi(t,\kappa)$.   We can in principle invert eq \ref{m33}
to obtain $\kappa$ in terms of $t$ and $\chi$: $\kappa =
K(t,\chi)$. 
This gives us the velocity function $\hat{V}(t,\chi) = \hat{V}^{*}(t,K(t,\chi).$
 If
the function $R(\kappa)$ were known, we would then also have $R(t,\chi)
= R(K(t,\chi))$.

$T(t,\chi)$
can be found by noting from eqs \ref{dc} that
\begin{equation}
T_\chi = -\frac{a\hat{V}}{c} T_t = -\frac{a\hat{V}}{c} \gamma = \kappa. \lbl{m34}\end{equation}
 
By substituting eq \ref{m34} into
eq \ref{20}, and integrating over $\kappa$ instead of $\chi$ by dividing the integrand by the partial of eq \ref{m33} with respect to $\kappa$, we find an expression for $T(t, \chi)$:
\begin{equation}
T(t,\chi) = t + \int_t^{\infty} \left[ 1 -
 \frac{1}{\sqrt{1 + \frac{c^2\kappa^2}{a^2}}} \right] \partial s_\kappa, \lbl{m35}\end{equation}
where $\kappa$ is put equal to $K(t,\chi)$ after integration at constant $\kappa$ in order to get $T(t,\chi)$. 

 This completes the solution.   
 Since $\kappa(R)$ can be any function that vanishes at the origin, there thus exist an infinite number of solutions for our transformed coordinates with $A=1, C=0$.  
 
\subsubsection{Independent determination of $c(t)$} \lbl{c(t)}

To determine physicality, we will next find $\frac{1}{R_\chi}$ close to the origin.  
$R_\chi = \frac{\kappa_\chi }{\kappa'(R)}$  can be written in an inverted form by taking the derivative of
eq \ref{m33} with respect to $\kappa$ at constant $t$:
\begin{equation}
 	\frac {1}{R_\chi} =  \kappa'\left[ \left( 	\frac{\partial K}{\partial \chi}\right)_{\!\! t} \right]^{-1} = \kappa'\left( \frac{\partial \chi}{\partial \kappa} \right)_t 
=  \kappa'(R)\int_t^{\infty}\frac{c^2\partial s_\kappa}{a^2(1 + \frac{c^2\kappa^2}{a^2})^{\frac{3}{2}}}. \lbl{m37}\end{equation}
By eq \ref{in2"}, the light speed is given by
\begin{equation} 
v_L = \frac{cA}{B} = \frac{c\gamma a}{R_\chi }= c\gamma a\kappa'(R)\int_t^{\infty}\frac{c^2\partial t_\kappa}{a^2(1 + \frac{c^2\kappa^2}{a^2})^{\frac{3}{2}}}
\lbl{in3"} \end{equation} 

To be physical $B = \frac{\gamma a}{R_\chi }\rightarrow 1$ as $R$ approaches $0$.  
Putting $\gamma = 1$,  $\kappa(0) = 0$, and $R_\chi = a$ 
in eq \ref{m37}, and changing the integration variable from $t$ to $a(t)$ gives
\begin{equation}
	\frac{1}{a} = \kappa' (0)  \int_a^\infty \frac{c da}{a^2 \dot{a}} ,
\lbl{m39} 
\end{equation}
remembering that the dot indicates differentiation by $\hat{t}$.  $\kappa'(0)$ is a constant to be determined by $c(t_0) = 1$.  Note that the integral of eq \ref{m39} is independent of the functional form of
$\kappa(R)$,  
and is therefor the same for all $\kappa(R)$.  It was Eq \ref{m37} that gave me the first indication that the light speed ($\frac{cR_\chi}{\gamma a}$) was variable, and that it was the same near the origin for all $\kappa(R)$.  

Eq \ref{m39} is an integral equation for $c(t)$.
By differentiation of both sides of eq \ref{m39} by $a$, we can obtain
\begin{equation}
	c(t) = \frac{1}{\kappa'(0)}\dot{a}, 
\lbl{m40}
\end{equation}
which, as we should expect, is the same $c(t)$ (see eq \ref{p10}) we showed for all physical coordinate systems for $\kappa'(0) = \frac{\dot{a}(t_0)}{c_0} = \frac{a_0 H_0}{c_0}$.  
This independent derivation of $c(t)$ confirms the validity of carrying the series expansion to second order since these complete transforms give the same $c(t)$.  

Notice that we have found this solution and the value for $c(t)$ without using the GR Field Equation nor any assumption about the variation of rest mass $m$ and gravitational constant $G$, just like the power series determination of $c(t)$.

\subsubsection{Zero density universe $\Omega = 0$}\lbl{pt.5}

It is interesting to consider the limiting case of a zero density universe: $\Omega = 0, \Omega_r = 1$, $a_0 H_0 = 1$ (eq  \ref{a4}).  Eq \ref{p10} makes $c = 1$.  Eq \ref{a7} makes $\dot{\alpha} = H_0$ for all $t,\chi$. Integrating gives $a = t$.  
Eq \ref{m33} gives $ \chi = \mbox{csch}^{-1}{\frac{t}{\kappa}}$, or $\kappa = K(t,\chi) = t \sinh{\chi}$. We can then find from eq \ref{m31'} that $V(t,\chi) = -\tanh{\chi}$ and from eq \ref{12'} that $\gamma = \cosh {\chi}$
so that 
\begin{equation}\label{a9'}
	c = 1 = \frac{R_\chi}{\gamma a} = \frac{dR}{d\kappa}\frac{K_{\chi}}{\gamma t}  = \frac{dR}{d\kappa} .
\end{equation}
Thus the physicality condition is met for all $R$ with $R = K$ and $A = 1, B=1$, so that
the complete transform with eq \ref{m35} becomes
\begin{equation}
\begin{array}{l}    R = t\sinh{\chi} ,\\ T = t \cosh{\chi}. \end{array} \lbl{51} \end{equation} 
These coordinates have been known ever since
Robertson \cite{R2} showed that this transformation from the FLRW co-moving
coordinates at zero density obeyed the Minkowski metric.
What is new is that this solution was derived from the equations we obtained for our physical time transforms with $A = 1$.  
It can also be obtained from the physical distance transforms ($B=1$) since eqs \ref{23} and \ref{r5}  for $\hat{V}$ become identical with $\hat{V}_t = 0$ and $\frac{a}{c} = a = t$.  It is the only known rigid physical coordinate system for all times and distances in a homogeneous and isotropic universe.  $R$ is plotted vs $T$ in Fig. 5 to show how similar it is to the physicality region of Figs. 2-4.\\\

\subsection{Physical Distance}\lbl{pd}

\subsubsection{Partial differential equation for $\hat{V}$.}

For diagonal coordinates with physical $dR$ for all $t$ and $\chi$, $B=1$, so eq \ref{in2} becomes
\begin{equation}
R_\chi = a\gamma
\lbl{r1}
\end{equation}
By integration we find
\begin{equation}
	R = a\int_0^\chi {\gamma \partial\chi_t},
\lbl{r2}
\end{equation}
and partial differentiation with respect to $t$ gives
\begin{equation}
	R_t = c\dot{a} \int_0^\chi{\gamma\partial\chi_t} +a\int_0^\chi{\gamma_t\partial\chi_t}.
\lbl{r3}
\end{equation}
We can then find $\hat{V}$ from eqs \ref{m13}, \ref{r1}, and \ref{r2} as
\begin{equation}
	\hat{V} = -\frac{R_t}{c\gamma} = -\frac{1}{c\gamma}\left[ c\dot{a} \int_0^\chi{\gamma\partial\chi_t} +a\int_0^\chi{\gamma_t\partial\chi_t}\right].
\lbl{r4}
\end{equation}
This is an integral equation for $\hat{V}$.  It can be converted into a partial differential equation by multiplying both sides by $\gamma$ and partial differentiating by $\chi$:
\begin{equation}
	\gamma^2 \left[ \hat{V}_x + \frac{a}{c} \hat{V} \hat{V}_t \right] = -\dot{a} = -\frac{1}{ c} \frac {da}{dt}.
\lbl{r5}
\end{equation}
Note that this is substantially different from the eq \ref{23} for $\hat{V}$ that we obtained for physical time. This means that it is not possible to find diagonal transforms with both physical time and physical distance for all values of $t$ and $\chi$ (except for $\Omega = 0$).  It is possible to have either one or the other be physical at all $t$ and $\chi$ with the other being physical only close to the origin.  

\subsubsection{General solution for $\hat{V}$}

Eq \ref{r5} can be solved as a standard initial-value problem.  Let $W = -\hat{V}$.  Eq \ref{r5} becomes
\begin{equation}
W_\chi - \frac{a}{c}WW_t = \frac{1}{c} \frac {da}{dt}(1-W^2)
\lbl{g5}\end{equation}
Define a characteristic for $W(t,\chi)$ by
\begin{equation}
(\frac{\partial t}{\partial \chi})_c = -\frac{a}{c}W
\lbl{g6}\end{equation}
so
\begin{equation}
(\frac{\partial W}{\partial \chi})_c = \frac{1}{c} \frac {da}{dt}(1-W^2)
\lbl{g7}\end{equation}
(The subscript $c$ here indicates differentiation along the characteristic).  If we divide eq \ref{g7} by eq \ref{g6} we get
\begin{equation}
(\frac{\partial W}{\partial t})_c = -\frac{1}{a} \frac {da}{dt}\frac{(1-W^2)}{W}
\lbl{g8}\end{equation}
This can be rearranged to give
\begin{equation}
\frac{W (\partial W) _c}{W^2-1} = \frac{(\partial a) _c}{a}
\lbl{g9}\end{equation}
This can be integrated along the characteristic with the boundary condition at $\chi=0$ that $W = 0$ and $a = a_c$:
\begin{equation}
1-W^2 = \frac{a^2}{a^2_c} = \frac{1}{\gamma ^2}.
\lbl{g10}\end{equation}
This value for $W$ (assumed positive for expanding universe) can be inserted into eq \ref{g6} to give
\begin{equation}
(\frac{\partial t}{\partial \chi})_c = -\frac{a}{c}\sqrt{1-\frac{a^2}{a^2_c}}
\lbl{g11}\end{equation}
We can convert this to a differential equation for $a$ by noting that $cd t_c = d \hat{t}_c = \frac{1}{\dot{a}} d a _c$
\begin{equation}
(\frac{\partial a}{\partial \chi})_c = -a \dot{a}\sqrt{1-\frac{a^2}{a^2_c}}.
\lbl{g12}\end{equation}
We can provide an integrand containing functions of only $\alpha$ by using the GR relation for $\dot{a}$ in eq \ref{a7}, which does not assume that $\dot{a} \propto c$. Eq \ref{g12} then becomes
\begin{equation}
(\frac{\partial \alpha}{\partial \chi})_c = -a_0 H_0 \alpha^2 E(\alpha)\sqrt{1-\frac{\alpha^2}{\alpha^2_c}}.
\lbl{g13}\end{equation}
This can be integrated along the characteristic with constant $\alpha_c$, starting with $\alpha = \alpha_c$ at $\chi= 0$.  This will give $\chi = X(\alpha,\alpha_c)$.  This can be inverted to obtain $\alpha_c(\alpha,\chi) $.  When this is inserted into eq \ref{g10}, we have a solution to eq \ref{g5} for $W(\alpha,\chi)$.  

I will now assume that $c \propto \dot{\alpha}$, then later show numerically that this makes $A \rightarrow 1$ as $R \rightarrow 0$ to prove physicality. (For $\Omega = 1$ in Sect \ref{fu1}, $c \propto \dot{\alpha}$ is shown explicitly).
Then $W(t,\chi)$ can be found from $W(\alpha,\chi)$ by using $\frac{c}{c_0} = \alpha E(\alpha)$ in eq \ref{p10} to get $t(\alpha)$:
\begin{equation}
t = \int_{0}^{\alpha}{\frac{d\alpha}{c\dot{\alpha}}} = \frac{1}{c_0 H_0}\int_{0}^{\alpha}{\frac{d\alpha}{\alpha^2 E^2}}. 
\lbl{g15}\end{equation}

\subsubsection{Obtaining $T,R$ from $\hat{V}$}

Eqs \ref{12}, \ref{m13}, and \ref{dc} show that
\begin{equation}
W = -\frac{a}{c}(\frac{\partial \chi}{\partial t})_R = \frac{a}{c}\frac{R_t}{R_\chi} = \frac{c}{a}\frac{T_\chi}{T_t}
\lbl{g16'}\end{equation}
so
\begin{equation} 
T_\chi - \frac {a}{c}WT_t = 0.
\end{equation}
Thus $T$ has the same characteristic as $W$ (eq \ref{g6}), so that $(\frac{\partial T}{\partial \chi})_c = 0$, and $T$ is constant along this 
characteristic:
\begin{equation}
T(t,\chi) = T(t_c,0) = t_c \equiv t(\alpha_c(t,\chi))
\lbl {g16}\end{equation}
where $t(\alpha)$ is given in eq \ref{g15} and $\alpha_c(\alpha(t),\chi)$ is found by inverting the integration of eq \ref{g13}.  This gives us the solution for $T(t,\chi)$ and $A$.  
\begin{equation} 
A = \frac{\gamma}{T_t} = \frac{a_c}{a}(\frac{\partial t}{\partial t_c})\chi = \frac{a_c}{a}\frac{\frac{d a_c}{d t_c}}{{\frac{d a}{dt}}} (\frac{\partial a}{\partial a_c}) _\chi .
\end{equation} 

The solution for $R$ can be obtained by integrating eq \ref{r2}, using $\gamma$ from eq \ref{g10} and $a_c(t,\chi)$ from eq \ref{g13}:
\begin{equation}
R(t,\chi) = a\int_{0}^{\chi}{\gamma  \partial \eta _t} = \int_{0}^{\chi}{a_c(t,\eta) \partial \eta _t} .
\lbl{g17}\end{equation}

Alternatively, for ease of numerical integration we would like to integrate $dR$ along the same characteristic as $T$ and $W$.  This can be obtained from the PDE
\begin{equation}                   
(\frac{\partial R}{\partial \chi})_c = R_\chi + R_t (\frac{\partial t}{\partial \chi})_c
\end{equation}
If we insert the values for these three quantities from eqs \ref{r1}, \ref{g16'}, and \ref{g6} , we get
\begin{equation}
(\frac{\partial R}{\partial \chi})_c = \gamma a + (\frac{cW}{a}) \gamma a(-\frac{aW}{c}) = \frac{a}{\gamma} = \frac{a^2}{a_c}.
\lbl{g19}\end{equation}

It is interesting that this solution for the physical distance coordinates (PD) is unique for each $a(t)$, whereas for the physical time coordinates (PT), there are an infinite number of solutions. This is because to obtain a solution for PD, we had to provide an additional relation, viz, for $\dot{a}$ (eq \ref{g13}), whereas for PT no additional relation was needed.
Possibly we could use the same relation in PT to make $\kappa(R) \propto R$ as for the similarity solution for a flat universe (Sect \ref{fu2}).  This would make PT unique as well, but I haven't been able to show this.

\subsection{Similarity solutions for flat universe, $\Omega = 1$}\lbl{fu}

I have found similarity integrations for the special case of $\Omega = 1$ where the GR solution is $a = a_0(\frac{t}{t_0})^{\frac{1}{2}}$ and $c = c_0(\frac{t_0}{t})^{\frac{1}{4}}$ (Sect \ref{c}).  
To simplify notation let us normalize time to $\frac{t}{t_0} \rightarrow t$, $\frac{a}{a_0} \rightarrow a$, and $
\chi \frac{a_0}{c_0t_0} \rightarrow x$, $\frac{T}{t_0} \rightarrow T $, $\frac{R}{c_0t_0} \rightarrow R$, and let $W = -\hat{V}$.

\subsubsection{Physical Distance}\lbl{fu1}

Eq \ref{r5} then becomes
\begin{equation}
	W_x - t^{\frac{3}{4}}WW_t = \frac{1}{2}t^{-\frac{1}{4}}(1-W^2)
\lbl{r6}
\end{equation}
This can be converted into an ordinary differential equation (ODE) by letting
\begin{equation}
	u = \frac{x}{t^{\frac{1}{4}}}
\lbl{r7}
\end{equation}
so that eq \ref{r6} becomes
\begin{equation}
	W'(1+\frac{uW}{4}) = \frac{1}{2}(1-W^2),
\lbl{r8}
\end{equation}
where the prime denotes differentiation by $u$.

Similarly we can find ODE's for $T$ and $R$ by defining:
\begin{equation}
	\frac{T}{t} \equiv q(u),	
\lbl{r9}
\end{equation}
and
\begin{equation}
	\frac{R}{t^{\frac{3}{4}}} \equiv s(u),
\lbl{r10}
\end{equation}
where $q(u)$ and $s(u)$, from eqs \ref{dc} and \ref{m13}, are given by the coupled ODE's:
\begin{equation}
	q'(1+\frac{uW}{4}) = qW,
\lbl{r11}
\end{equation}
and
\begin{equation}
	s'(W+\frac{u}{4}) = \frac{3}{4} s
\lbl{r12}
\end{equation}
It is useful to find that $q = \gamma^2$, $s' = \gamma$, $s = \gamma \frac{u+4W}{3}$, and $A = \frac{\gamma}{T_t} = \frac{1+\frac{uW}{4}}{\gamma} = \frac{v_p}{c} $; so $T = \gamma^2 t$, and $R = t^{\frac{3}{4}}\gamma \frac{u+4W}{3}$.

For small values of $u$, $W = \frac{u}{2}$, $q = 1 + \frac{u^2}{8}$, $s = u$, $\frac{v_p}{c} = 1+\bigcirc{(W^4)}$, and $R = t^{\frac{1}{2}}x = ax$.  The light speed $v_p$ measured on AP remains close to that measured on FLRW out to large $R$.  We also note that $T_t \rightarrow 1 + \frac{W^2}{2}$, confirming that these coordinates have physical time close to the origin, justifying $c(t) = t^{-\frac{1}{4}} $.  

An alternate approach would be to start with $c(t)$ unknown, but of the form $c = t^{-b} $.  Then the GR Field Equation eq \ref{f6}  will give $\alpha = t^d$, where $d = \frac{2}{3}(1-b)$. Eq \ref{r6} then becomes $W'(2+uWd) = 2d(1-W^2)$, where the independent variable is $u = \frac{x}{t^{\frac{d}{2}}}$.  For $T = tq$ this will make $q'(2 + uWd) = 2qW$ and $T_t = q - \frac{udq'}{2}$.  For small $u$, $q \rightarrow 1+\frac{d}{2}u^2$, $W \rightarrow ud$, and $T_t \rightarrow 1 + \frac{(1/d - 1)}{2}W^2$.  To be Lorentz  $T_t \rightarrow 1+\frac{W^2}{2}$ so that $d = \frac{1}{2}$ and $b = \frac{1}{4}$, confirming that $c \propto \sqrt{\frac{da}{dt}}$.  

For constant light speed, $b = 0, d = \frac{2}{3}$ and $T_t \rightarrow 1 + \frac{W^2}{4}$, slower than Lorentz as found in Fletcher\cite{F2}.  This has implications for the use of the GR Field Equation.  We can't integrate the Physical Distance transforms without using the FE.  When we use it for a constant light speed, we don't get the Lorentz transform for small $\hat{V}$.  When we use it for an arbitrary varying light speed, we get the Lorentz transform when we use the same $c(t)$ as for the power series and for the Physical Distance transform.  This self-consistency indicates
 that we are using the correct FE and the correct $c(t)$.

As $t \rightarrow 0$, $u \rightarrow \infty$, $\gamma \rightarrow \kappa u^2$, $W \rightarrow 1 - \frac{1}{2\kappa^2 u^4} $, $q \rightarrow \kappa^{2}u^4 $, and $s \rightarrow  \kappa \frac{u^3}{3}$.  $T$ and $R$ both remain finite at this limit with $T \rightarrow \kappa^{2} x^4 $, and $R \rightarrow  \kappa \frac{x^3}{3}$, where $x_L \rightarrow 4$ at $t \rightarrow 0$.  $\kappa$ is difficult to determine from the numerical integration because of the singularity at large $u$, but my integrater gives .0364.   
The fact that $T$ does not go to zero when $t$ goes to zero results from equating $T$ with $t$ at $t = 1$ and not at $t = 0$. 

The distance $R$ and time $T$ can be found from the numerical integration of the coupled ODE's. 
The paths of galactic points are those for constant $x$.  The path photons have taken reaching the origin at $t_1$ is found by calculating $x_p$ vs $t$ and using the transform to $T,R$.  Thus, for $\Omega = 1$
\begin{equation}
	x_p = \int_{t_1}^t{\frac{c}{a} dt} = 4(t_1^{\frac{1}{4}}-t^{\frac{1}{4}} )
\lbl{r13}
\end{equation}
For light arriving now, $t_1 =1$, the value of $u_p$ becomes
\begin{equation}
	u_p = 4(\frac{1}{t^{\frac{1}{4}}} -1) = 4(c - 1),
\lbl{r13'} \end{equation}
where we inserted $c = \frac{1}{t^{\frac{1}{4}}}$ to obtain the relation of $c$ to $u_p$.  

Galactic and photon paths are shown in Fig. 1 and 2.  An approximate upper limit of physicality is shown by the heavy dotted line: $A = .953$, $\frac{W^2}{2} = 0.253$, $u = 2.0$, $R = 2.30 t^{\frac{3}{4}} = 1.35T^{\frac{3}{4}}$.  
$R$ vs $T$ at $t=0$ provides a non-physical horizon: $R_h = 1.747T_h^{\frac{3}{4}}$.  

It is also interesting to calculate the acceleration $g$.  If we insert the values of $V$, $R$, and $\frac{a}{c}$ in eq \ref{g}, we obtain
\begin{equation}
	-g = \frac{1}{8\gamma t}\left[ \frac{u}{1+\frac{uW}{4}} \right],
\lbl{gB} \end{equation}
where the units of $g$ are $\frac{c_0}{t_0}$.
For small $u$ close to $t = t_0$, $g$ goes to zero as $-\frac{u}{8}$.  

Since small $u$ is the region with physical coordinates, it is interesting to express $g$ in unnormalized coordinates:
\begin{equation}
	-g = \frac{1}{8} \frac{c_0}{t_0} \frac{R}{c_o t_0} = G_0 \rho_0 \frac{4\pi}{3}R,
\lbl{gB1}\end{equation}
where we have used $\Omega = 1$ in eq \ref{a3}.  For small $t$, $-g$ goes to $\infty$ as $\frac{1}{(2\gamma t^{\frac{5}{4}}}) = 1.2 t^{-\frac{3}{4}}$ along the light path.    
At the physicality limit, $-g = .69 \frac{c_0}{t_0} = 12.5 \times 10^{-9} $$\frac{m}{sec^2}$.  

$g$ can be obtained from a gravitational potential using $g = -\frac{d\phi}{dR}$, which for close distances is:
\begin{equation}
	\phi = G_0 \rho_0 \frac{2\pi}{3}R^2 = \frac{R^2}{16t_0^2} = c_0^2\frac{u^2}{16}.
\end{equation}

 The slope of the light path in Fig. 1, a coordinate velocity of light, can be shown in normalized units for this incoming light path to be
\begin{equation}
	v_L = (\frac{dR}{dt})_{L} =  -(1+ \frac{u}{4}) \left[ \frac{1-W}{1+W} \right]^{\frac{1}{2}}.
\lbl{vB2}\end{equation}
For small $u$ , 

\begin{equation}
	v_L \rightarrow \\1 - \frac{u}{4}.  
\lbl{vg}\end{equation}
For the outgoing light path
\begin{equation}
	v_L = (\frac{dR}{dt})_{L} =  (1- \frac{u}{4}) \left[ \frac{1+W}{1-W} \right]^{\frac{1}{2}}
\lbl{vB2'}\end{equation}
For small  $u$,  
\begin{equation}
	v_L \rightarrow 1 + \frac{u}{4}  ,
\lbl{vg'}\end{equation}
Thus, the coordinate light speed has a different $u$ dependence on $R$ for incoming and outgoing light paths because the slope is dependent on $t$, not $R$.  This differs from the Schwarzschild solution that has the same $R$ dependence of the coordinate light speed for both directions of the light path.
 
The observed light at the origin $\nu$ that is emitted from a AP source at $R$ as $\nu_0$ 
is also smaller than the same light emitted at the origin.  This can be shown to be
\begin{equation}
	\frac{\nu}{\nu_0}  = (\frac{\partial T}{\partial t})_R = \frac {1}{\gamma A} = q \frac{1-W^2}{1 + \frac{uW}{4}}
\end{equation}
Close to the origin it is:
\begin{equation}
	\frac{\nu}{\nu_0} = 1 - \frac{\hat{V}^2}{2} = 1 - \frac{u^2}{8} = 1 - \frac{2\phi}{c_o^2}.
\lbl{vB3}\end{equation}
This, of course, is the same as a dilation effect for a collocated galactic point at $R$ that shows up as a gravitation red shift at the origin due to homogeneity of $t$.
  
\subsubsection{Physical Time}\lbl{fu2}

There is also a similarity solution for physical time, $A=1$, for $\Omega = 1$.  With the same normalizations as above, using eq \ref{23}, the ODE for $W$ is
\begin{equation}
	W'(W+\frac{u}{4}) = \frac{3}{4}W(1-W^2),
\lbl{vA1} \end{equation}
with the ODE's for $T,R,q,s,x_p)$ the same.  This is the same as physical distance for small $u$, but differs numerically at large $u$.  Useful relations for physical time are obtained from the general solution in Appendix \ref{pt}: $R = 2\kappa$, $q = \gamma (1+\frac{uW}{4})$, $s = 2\gamma W$, and $B = \gamma\frac{u+4W}{3s} = \frac{2}{3}(1+\frac{u}{4W})$.  These can be used to find the gravitational field from eq \ref{g}:
\begin{equation}
	g = -\frac{3}{2t^{\frac{5}{4}}}\frac{\gamma W^2}{u+4W}
\lbl{vA2} \end{equation}
and the coordinate velocity for an incoming light path:
\begin{equation}
	v_L = -\frac {3}{2}  \left[ \frac{(1+\frac{u}{4})}{1+\frac{u}{4W}} \right] \left[ \frac{1-W}{1+W} \right] ^{\frac{1}{2}}
\lbl{vA3} \end{equation}
Eqs \ref{vA2} and \ref{vA3} approach the same values as physical distance for small $u$.

\section{Gravitational field in the FLRW and AP coordinates} \lbl{s2.1.1}

We wish to find the components of the radial acceleration of a test particle located at R in the AP transformed system.  We will do this by calculating the FLRW components of the acceleration vector and find the transformed components by using the known diagonal transforms.  For the FLRW components, we will use the metric
\begin{equation}
	ds^2 = d\hat{t}^2 - a^2 d\chi^2 -a^2r^2 d\theta ^2 -a^2r^2\sin^2{\theta} d\phi^2 .
\end{equation}
 
Let
\begin{equation}
	x^1 = \chi,~ x^2 = \theta,~ x^3~ = \phi,~x^4 = \hat{t}, \lbl{m1}\end{equation}
and the corresponding metric coefficients become
\begin{equation}
g_{44} = 1,~g_{11} = -a^2,~g_{22} = -a^2r^2,~g_{33} =
-a^2r^2\sin^2{\theta}. \lbl{m2}\end{equation}

For any metric, the acceleration vector for a test particle is 
\begin{equation}
	A^{\lambda} = \frac{dU^{\lambda}}{ds} + \Gamma_{\mu \nu}^{\lambda} U^{\mu} U^{\nu}, 
\lbl{m3}\end{equation}
where the $\Gamma$'s are the affine connections and  $U^\lambda$ is the velocity vector of the test particle.  In our case the test particle is at the point $R$ on the transformed coordinate, but not attached to the frame so that it can acquire an acceleration. Instantaneously, it will have the same velocity as the point on the transformed coordinate, and its velocity and acceleration vectors will therefor transform the same as the point (eq \ref{11}).

We will be considering accelerations only in the radial direction so that we need find
affine connections only for indices 1,4.  The only non-zero
partial derivative with these indices is
\begin{equation}
\frac{\partial g_{11}}{\partial x^4} = -2a\dot{a}. \lbl{m5}\end{equation}
 
The general expression for an affine connection for a diagonal metric is
\begin{equation}
 \Gamma^\lambda_{\mu\nu}   = \frac{1}{2g_{\lambda\lambda}}\left[ \frac{\partial g_{\lambda\mu}}{\partial
 x^\nu} + \frac{\partial g_{\lambda\nu}}{\partial
 x^\mu} - \frac{\partial g_{\mu\nu}}{\partial
 x^\lambda}\right]. \lbl{m6}\end{equation}
The only three non-zero affine connections with 1,4 indices
are
\begin{equation}
\Gamma^4_{11} = a\dot{a},~\Gamma^1_{41} = \Gamma^1_{14} =
\frac{\dot{a}}{a}. \lbl{m7}\end{equation}
The acceleration vector in FLRW 
coordinates of our test particle moving at the same velocity as a point on the transformed frame becomes 
\begin{equation}
\begin{array}{l}
A^{\hat{t}} = \frac{dU^4}{ds} + \Gamma^4_{11}U^1 U^1, \\[2mm]
A^\chi = \frac{dU^1}{ds} + \Gamma^1_{41}(U^4 U^1 + U^1 U^4),  \end{array} \lbl{m8}\end{equation}
Using $U^4$ and $U^1$ in eq \ref{u2} we find
\begin{equation}
\begin{array}{l}
A^{\hat{t}} = \gamma\left( \frac{\partial \gamma}{\partial \hat{t}}\right)_{\! R}
 + a\dot{a}\frac{\gamma^2\hat{V}^2}{a^2} = \gamma^4\hat{V}\left( \frac{\partial \hat{V}}{\partial \hat{t}}\right)_{\! R}
 + \frac{\dot{a}}{a}\gamma^2\hat{V}^2,\\[2mm]
A^\chi  = \gamma
\left( \frac{\partial}{\partial \hat{t}}(\frac{\gamma\hat{V}}{a}) \right)_{\! R}
+ 2\frac{\dot{a}}{a}\frac{\gamma^2\hat{V}}{a} = \frac{\gamma^4}{a}   \left( \frac{\partial \hat{V}}{\partial \hat{t}}\right)_{\! R} + 
\frac{\dot{a}}{a^2}\gamma^2\hat{V}. \end{array} \lbl{m9}\end{equation}
Since the acceleration vector of the test particle at $R$ in the transformed coordinates will be orthogonal to the velocity vector, it becomes
\begin{equation}
\begin{array}{l}
A^T = 0,\\[2mm]
A^R = \frac{DU^R}{Ds}  \equiv -\frac{g}{c^2}. \end{array} \lbl{m23}\end{equation}
$A^R$ is the acceleration of a point on the $R$ axis, so the gravitational field affecting objects like the galactic points is the negative of this.  
$g$ is defined so that $mg$ is the force acting on an object whose mass is $m$. 
For a range of time in which $c(t)$ is reasonably constant, $g = \frac{d^2R}{dT^2}$, the normal acceleration.  Since the vector $A^\lambda$ will transform like $dT,dR$ (eq \ref{7}):
\begin{equation}
	A^R = \frac{1}{c}R_tA^{\hat{t}} + R_\chi A^\chi  
\end{equation}
so that
\begin{equation}
	-\frac{g}{c^2} = \left[ \gamma^4 \hat{V}\left( \frac{\partial \hat{V}}{\partial \hat{t}}\right)_{\! R} +
\frac{\dot{a}}{a}\gamma^2\hat{V}^2 \right] \frac{1}{c}R_t + \left[  \frac{\gamma^4}{a}\left( \frac{\partial \hat{V}}{\partial \hat{t}}\right)_{\! R} +
\frac{\dot{a}}{a^2}\gamma^2\hat{V} \right] R_\chi 
\end{equation}

With the use of eq \ref{m13}, this can be simplified to
\begin{equation}
	-\frac{g}{c^2} = \frac{R_\chi}{a} \left[ \gamma^2 \left( \frac{\partial \hat{V}}{\partial \hat{t}}\right)_{\! R} + \frac{\dot{a}}{a} \hat{V} \right] 
\lbl{g} \end{equation}

In terms of the normalized coordinates for a flat universe (Appendix \ref{fu}), this becomes
\begin{equation}
g = -\frac{s'}{t}\left[ \gamma^2W'(\frac{u}{4}+W)-\frac{W}{2} \right]
\lbl{fug}\end{equation}
The acceleration $g$ can be thought of as the gravitational field caused by the mass of the surrounding galactic points, which balances to zero at the origin, where the frame is inertial, but goes to infinity at the horizon.  It is the field which is slowing down the galactic points (for $\Lambda = 0$).  It is also the field that can be thought of as causing the gravitational red shift (App. \ref{fu}). 

\section{Special and General relativity extended to include a variable light speed} \lbl{s2.1.2}

\subsection{ Introduction} \lbl{intro}

The aim of this section is to outline a way that not only Lorentz, but all of special (SR) and general relativity (GR) can be extended to allow a variable light speed with minimal changes from standard theory.
 The extended Lorentz transform for local coordinates is derived from the basic assumption of relativity that the light speed $c$ is the same for all moving observers at the same space-time point even though the light speed and their relative velocity $V$ may vary.  To form SR vectors and tensors we use a differential construct 
$d\hat{T} = cdT$ from physical time $T$\cite{Mag3} and a dimensionless velocity $\hat{V} = \frac{V}{c}$.  
In addition we propose that the rest mass of a particle varies so as to keep its rest energy constant.  This seems reasonable in order to eliminate the need for an external source or sink of energy for the rest mass.
These  assumptions simplify the construction of SR vectors and conserves the stress-energy tensor of an ideal fluid.  For GR, we propose the standard GR Action, but use the extended stress-energy tensor and allow the gravitational constant $G$ to vary with $c$.  The variable light speed is introduced in the line element that determines the space-time curvature.
 
We will use the notation $t$ for time when the light speed is $c(t)$, as it must be for a uniform and isotropic universe if it is to be variable.  Then $\hat{t}$ can be a transform from $t$: $\hat{t} = \int{c(t)dt}$. 
The GR curvature tensor is derived from a line element that typically has the time $t$ appearing in the combination of $c(t)dt$ that would require the tensor to contain the derivatives of $c$.  
The use of $\hat{t}$ instead of $t$ eliminates these derivatives without changing the relations of the components of the tensors, and also allows all the relations of curvelinear coordinates used for constant $c = 1$ to be retained.    
Then, from a solution with $\hat{t}$ the observable physical $t$ can be found with a transform from $\hat{t}$ to $t$.

\subsection{The extended Lorentz transform and Minkowski metric}\lbl{LT}

Let us consider two physical frames moving with respect to each other.  The first frame $(S)$ will have clocks and rulers whose readings we will represent by $T$ and $x$.  The second frame $(S^*)$ will move in the $x$ direction at a velocity of $V = (\frac{\partial x}{\partial T})_{x^*}$ as measured by $T$ and $x$ and will  have clocks and rulers whose coordinates we will represent by $T^*$ and $x^*$.  The velocity of the first frame will be $V^*$ as measured by $T^*$ and $x^*$.  We assume that the light speed, even though variable, is the same as measured on both frames at the same space-time point.  We also allow $V$ to be variable.

 In order for $S$ to measure the small separation of points $\Delta x^*$ on $S^*$, $S^*$ sends two simultaneous ($ \Delta T^* = 0$) signals as measured on its clocks, one at the beginning of $\Delta x^*$ and the other at the end.  $S$ measures the space between the signals as $\Delta x$, but does not see these signals as simultaneous.  The far end signal is delayed by $\Delta T$ over the near end signal for this reason.  $S$ measures $\Delta x^*$ to be the distance $\Delta x$ reduced by the distance that $S^*$ has traveled in the time $\Delta T$ after $S$'s simultaneity ($\Delta T = 0 $) with the near end, i.e., $\Delta x - V \Delta T$.  Since we are looking for linear relationships, we assume that the $S^*$ measure of $\Delta x^*$ is proportional to the $S$ measure:
\begin{equation} 
\Delta x^* = \alpha (\Delta x - V \Delta T)
\lbl{lt2} \end{equation}
where we have allowed $\alpha$ and $V$ to be varying, but approach a constant value for small $\Delta$ 's.
  We also assume that for the two Cartesian directions $\Delta y$ and $\Delta z$ perpendicular to the motion along $x$ that the $S^*$ and $S$ coordinates are the same 
\begin{equation} \Delta y = \Delta y^*, \Delta z = \Delta z^*
\lbl{lt1}\end{equation}  
and that the time $T$ does not depend on $y$ or $z$.  $\alpha$ will be determined from the assumption that the light speed is the same on all moving frames. We will adapt the analysis of Bergmann\cite[pp33-36]{brg} to a variable light speed.  Choosing the point of origin so that $\Delta T$ and $\Delta T^*$ vanish when $\Delta x$ and $\Delta x^*$ vanish, we expect that $\Delta T^*$ will be a linear function of $\Delta T$ and $\Delta x$: 
 
\begin{equation} 
\Delta T^* = \gamma \Delta T + \zeta \Delta x
\end{equation} 
where $\alpha, \gamma$, and $\zeta$ are slowly varying functions that approach a constant for small $\Delta$'s.  We will now determine their values.

We assume that the light speed can be variable, but in small intervals of time and distance it will be almost constant.  It will
have the same values in $S^*$ as in $S$ at the same space-time point.  For light moving in an arbitrary direction, each measures the light speed $c$ as the change in distance divided by the change in time of its own coordinates:
\begin{align}
\Delta x ^2 + \Delta y ^2 + \Delta z^2 & = c^2  \Delta T^2, \lbl{lt3'}\\
\Delta x{^*} ^2 + \Delta y{^*} ^2 + \Delta z{^*}^2 & = c^2  \Delta T{^*}^2, \lbl{lt3}
\end{align}
where we have chosen an origin where all the $\Delta$'s vanish.  
 By using Eqs \ref{lt1} and \ref{lt2} in Eq \ref{lt3}, we can eliminate the starred items to get
\begin{equation} 
 \alpha ^2 (\Delta x - V \Delta T)^2 + \Delta y^2 + \Delta z^2 = c^2(\gamma \Delta T + \zeta\Delta x)^2. 
\end{equation}
We can rearrange the terms to obtain
\begin{equation} 
 (\alpha ^2 - c^2\zeta ^2)\Delta x^2 - 2(V\alpha ^2+c^2 \gamma \zeta) \Delta x \Delta T + \Delta y^2 +\Delta z^2 = (c^2\gamma ^2 -V^2 \alpha ^2)\Delta T^2.
\end{equation} 
If we compare this to eq \ref{lt3'} we get
\begin{align}
c^2 \gamma ^2 -V^2 \alpha ^2 & = c^2 \\
\alpha ^2 - c^2 \zeta ^2 & =1 \\
V \alpha ^2 + c^2 \gamma \zeta & = 0
\end{align}
We can solve these three equations for the three unknowns $\alpha, \gamma$, and $\zeta$:
\begin{align}
\gamma^2 & = \frac {1}{1-\frac{V^2}{c^2}} \\
\zeta & = \frac {1-\gamma^2}{\gamma V} =- \frac {\gamma V}{c^2} \\
\alpha ^2 & = -\frac{c^2 \gamma \zeta}{V} = \gamma^2
\end{align}

Thus in the differential limit of $\Delta 's$ going to zero, we write them as differentials, so the relation of differentials becomes 
\begin{align}
dT^* & = \gamma(dT-\frac{V}{c^2}dx), \lbl{lt} \\
dx^* & = \gamma(dx - VdT),
\end{align} 
 By inverting this we get
\begin{align}
dT & = \gamma(dT^*+\frac{V}{c^2}dx^*), \lbl{SR} \\
dx & = \gamma(dx^* + VdT^*). \lbl{sr'}
\end{align}
so $V{^*} = -V$ as you would expect.

This is the same as for a constant $c$, except here $c$ has been allowed to vary.  

We define a line element $ds$ by the relation
\begin{equation} 
ds^2 \equiv c^2 dT^2 - dx^2 - dy^2 - dz^2
\lbl{m1}\end{equation}
If we substitute eqs \ref{lt1}, \ref{SR} and \ref{sr'} into eq \ref{m1}, the form is the same:
\begin{equation} 
ds^2 = c^2 dT{^*}^2 - dx{^*}^2 - dy{^*}^2 - dz{^*}^2
\lbl{m2} \end{equation}
That is, the extended world line is invariant in form to changes in coordinates on frames moving at different velocities. The line element is symmetric in the spatial coordinates, so it is valid for motion in any direction.  In polar coordinates this becomes
\begin{align} 
ds^2 & = c^2dT^2 - dR^2 - R^2 d\theta^2 -R^2 sin^2\theta d \phi ^2 \lbl{sr3'} 
  \end{align}
This is the Minkowski line element ($\hat{M}$) extended to allow for a variable light speed.  Both $\hat{L}$ and $\hat{M}$ are valid in the two dimensions $T$ and $R$ even if the metrics of $S^*$ and $S$ did not have equal transverse differentials, but had no transverse events ($d\theta = d\phi = d\theta^* = d\phi^* = 0$).

Notice that if we divide eqs \ref{m1} and \ref{m2} by $c^2$  the two equations still have identical forms, so that the differential time $ d \tau \equiv \frac{ds}{c}$ is also invariant in form to $\hat{L}$ transforms.  Since $d\tau = dT$ for constant spatial coordinates, $\tau $ is the time on a clock moving with the frame.  

This derivation has depended on a physical visualization so that we assume that differentials that represent physical time and radial distance must have a $\hat{M}$ metric for their time and distance differentials in at least two dimensions and an extended Lorentz transform $\hat{L}$ to other collocated physical differentials of time and distance on a frame moving at a velocity $V$.  We will call such differentials physical coordinates.  Time and distance coordinates that do not have these relations will not be physical; one or the other may be physical, but not both unless they have a $\hat{M}$ metric.

The extended Lorentz transform $\hat{L}$  can be written in a symmetric form using $d\hat{T} \equiv cdT$ and $\hat{V} \equiv \frac{V}{c}$ with the velocity in the $R$ direction (as it will be in a homogeneous and isotropic (FLRW) universe):
\begin{align}
d\hat{T}^* & = \gamma(+d\hat{T} - \hat{V} dR), \lbl{lt5}\\
dR^* & = \gamma(-\hat{V} d\hat{T} +dR).
\end{align}

In general for a varying $c$, $\hat{T}$ is not a transform from $T$ alone, although, as we have shown in eq \ref{lt5}, we can use the construct $d\hat{T} = cdT$ to describe the $\hat{L}$ transform.
In a FLRW universe for events in the radial direction measured by the variables $(t,\chi)$, if $c$ is variable, it is a simple function of $t$ since homogeneity in space makes it independent of $\chi$. In this case $\hat{t}$ is a transform from $t$ alone (e.g., eq \ref{3'}).  

\subsection{Extended SR particle kinematics using contravariant vectors}\lbl{sr}

In this section I will outline the way vectors and tensors can be defined when the light speed is variable.   
In Cartesian coordinates, let $dx^1,dx^2,dx^3 = dx,dy,dz $, and $dx^4 = d\hat{T} = cdT$.  The $\hat{M}$ metric then becomes
\begin{equation}
	ds^2 = \eta_{\mu \nu}dx^\mu dx^\nu,
\lbl{sr1"} \end{equation}
where $\eta_{\mu \nu} = (-1,-1,-1,+1)$ for $\mu = \nu$, and zero for $\mu \ne \nu$. The velocity $\dot{x}^\mu$ is $\frac{dx^\mu}{d\hat{T}} = \frac{V^\mu }{c}$ with $\dot{x}^4 = 1$.  (The dot represents the derivative with respect to $d\hat{T}$).  The world velocity becomes 
\begin{equation} 
{U}^\mu = \frac{dx^\mu}{ds} = \gamma \dot{x}^\mu.
\end{equation}
  $\dot{x}^\mu$ and $U^\mu$ are therefor dimensionless.  In order to make the rest mass energy constant, we define $\hat{m} = mc^2$ and the extended energy-momentum vector as 
\begin{equation}
	P^\mu = \hat{m}U^\mu = \hat{m} \gamma \dot{x}^\mu, 
\lbl{sr2}\end{equation}
so that $P^4 = \hat{m} \gamma = E$, the particle energy.  If $p$ is the magnitude of the physical momentum ($\gamma mV$), the EP vector magnitude is $E^2 - c^2 p^2 = \hat{m}^2 $.  It has units of energy rather than momentum or mass.

The $\hat{L}$ transform for the components of the EP vector is
\begin{equation}
	E^* = \gamma(E- \hat{V} pc).
\lbl{sr3} \end{equation}
For photons, $\hat{m} = 0$, so $E = h \nu$ and $p = \frac{h}{\lambda }= \frac{h\nu }{c}$, and the $\hat{L}$ transform is 
\begin{equation}
	\nu^* = \gamma \nu (1-\hat{V}).
\lbl{sr4} \end{equation}
This is the familiar relativistic Doppler effect.  

The force vector becomes 
\begin{equation}
	F^{\mu} = \frac {dP^{\mu}}{ds} = \hat{m} A^{\mu} = \hat{m} \frac {dU^{\mu}}{ds} .
\lbl{sr5'}\end{equation}
The first three components $\frac{F^i}{\gamma}$ will
 be the force $f^i$ felt by an object of mass $m$ when the light speed is $c$ ($i$ represent the three spatial coordinates).  In taking the derivative of $P^i$, we are implying that
$mc\frac{d(\frac{\gamma V}{c})}{dT}$ is more fundamental in determining the physical force than $m \frac {d(\gamma V)}{dT}$ when the light speed is variable.
We can express the gravitation force in the usual way as $mg^i$, where $g^{i} = A^{i}\frac{c^2}{\gamma}$.  $\frac{c}{\gamma}F^4$ is the rate of work $f^i V^i$ required to change the rate of change of energy $\frac{d(\gamma \hat{m})}{dT}$.
All these world vectors are invariant to the $\hat{L}$ transform and the $\hat{M}$ line element.  They become the usual vectors when $c$ is constant.

\subsection{Extended analytical mechanics}

We will next show how the Euler-Langrange equations apply to extended particle kinematics\cite{brg}.  For a mechanical system with conservative forces in (n+1)-dimensional space whose differentials are
$(dx^i,d\hat{T})$, the action $S$ is
\begin{equation} 
S = \int{L_p}ds.
\end{equation} 
Minimizing $S$ gives relations for $L_p$, the particle Lagrangian.   With no force acting, we will use
 
 \begin{equation} 
L_p = \hat{m}\sqrt{\eta _{\mu \nu}U^\mu U^\nu},
\end{equation}
so the momenta are
\begin{equation} 
P_\mu = \frac {\partial L_p}{\partial U^\mu} = \frac {\hat{m} \eta_{\mu \nu} U^\nu }{\sqrt{\eta _{\mu \nu}U^\mu U^\nu}}.
\end{equation} 
The root in this equation has the value 1 which makes it possible to solve it for $U^\mu $
\begin{equation} 
U^\mu =\frac { \eta ^{\mu \nu} P_\nu }{\hat{m}\sqrt{\eta _{\mu \nu} U^\nu U^\nu}} = \frac {P^\mu}{\hat{m}}, 
\end{equation} 
consistent with eq \ref{sr2}.

So,
\begin{equation} 
U^\mu P_\mu = \frac{\eta^{\mu \nu}P_\mu P_\nu}{\hat{m}}.
\end{equation} 
The Hamiltonian  $\hat{H}$ becomes
\begin{equation} 
\hat{H} = -L_p + U^\mu P_\mu = -\sqrt{\eta^{\mu \nu}P_\mu P_\nu } + \frac{\eta^{\mu \nu}P_\mu P_\nu}{\hat{m}}. 
\end{equation} 
Let $p \equiv \sqrt{\eta^{\mu \nu}P_\mu P_\nu }  = \hat{m}$, so 
\begin{equation} 
\hat{H} =  \frac {p^2}{\hat{m}} - p
\end{equation} 
Thus $\hat{H}$ vanishes,  but its derivative with respect to $P_\mu$ does not:
\begin{align}
U^\mu & = \frac {\partial \hat{H}}{\partial P_\mu} = 2\frac{\eta^{\mu \nu}P_\mu } {\hat{m}}
- \frac{\eta^{\mu \nu}P_\mu } {p}=  \frac {P^\mu}{\hat{m}} \\
\frac {dP_u}{ds} & = -\frac {\partial \hat{H}}{\partial x^\mu} = 0,
\end{align}
$P_\mu$ is conserved since we have considered no force acting.

\subsection{Extended stress-energy tensor for ideal fluid}\lbl{gr}

An ideal fluid can be treated in a similar way.  It is a collection of $n$ particles per unit volume of mass $m$.  We can form a rest energy density function $\hat\rho = n\hat{m}$.  In this case, $\hat{\rho}$ is not constant because $n$ is a a function of time and distance.  
We will use $t$ instead of $T$ to indicate that we are initially limiting this analysis to a rest frame of FLRW attached to a galactic point where $c$ is a function of $t$.  This can be transformed to other frames by a $\hat{L}$ transform.  It turns out that $\hat{\rho}$ using $d\hat{t}$ and $u^\mu = \frac{V^\mu }{c(t)}$ has much the same properties as $\rho  = nm$ using $dt$ and $V^\mu$ with constant $c$.  

The conservation law for particles in nonrelativistic terms for $n$ flowing at a velocity $V^i = cu^i$ is
\begin{equation} 
\frac { \partial n }{\partial \hat{t}} +  n u^i_{,i} = 0.
\end{equation}
where we have assumed that the differential of $c$ with distance is zero. 
For the conservation of energy we must include the stress forces $t^{ij}dA_j$ operating on the area of the differential volume, like the pressure $p$ where $t^{ij} = p\delta ^{ij}$.  We can convert the area stress forces by Gauss' theorem to a volume change in momentum to give a total 3D energy flux of $cP^i$, where
\begin{equation} 
P^i = \hat{\rho}u^i + u^j t^{ji}.
\end{equation}
The conservation of the fluid rest energy ($u^i = 0$) then becomes
\begin{equation} 
\frac{ \partial \hat {\rho }}{\partial t} + \mbox {div} (cP^i) = 0,
\end{equation}
 or
\begin{equation} 
\frac{ \partial \hat {\rho }}{\partial \hat{t}} + P^i_{,i} = 0.
\lbl{if2} \end{equation}

The Newtonian law linking the rate of change of the generalized velocity $u^i = \frac {dx^i}{d\hat{T}}$ to the force per unit volume $f^i$ in nonrelativistic terms can be written as
\begin{equation} 
\hat{\rho} \frac {d u^i}{d\hat{t}} = f^i 
\end{equation}
We can follow through the steps in any of the standard texts \cite{brg} to obtain the generalized
 stress-energy tensor of an ideal fluid in its rest frame to be
 \[
	 T^{\mu \nu} = T_{\mu \nu} =
	\begin{pmatrix}
		& p
		& 0
		& 0
		& 0 \\[10pt]
		& 0
		& p
		& 0
		& 0 \\[10pt]	
		& 0
		& 0
		& p
		& 0 \\[10pt]
		& 0
		& 0
		& 0
		& \hat{\rho}
 \end{pmatrix}.\] \lbl{if4} 
This is used in Sect \ref{FE}.

This can be generalized for a frame moving at a world velocity $U^\mu$:
\begin{equation}
	 T^{\mu \nu} = (\hat{\rho} + p)  U^\mu U^\nu  - p\eta^{\mu \nu}.
\lbl{sr5}\end{equation}
One can see that this is the same tensor since in the rest frame of the fluid $\hat{T} = \hat{t}$, $U^i = 0$, $U^4 = 1$.

The divergence of the stress-energy tensor is the force per unit volume:
\begin{equation} 
 T^{\mu \nu}_ {\quad,\nu} = F^\mu
\end{equation} 

The conservation of rest energy density (eq\ref{if2}) can then be written:
\begin{equation} 
 F^\mu =  T^{\mu \nu}_{\quad,\nu}  = 0.
\end{equation}

\subsection{Extended electromagnetic vectors and tensors}\lbl{em}

We will assume that the light speed that appears in electromagnetic theory (E/M) is the same as appears in relativity theory.  If it were not so, it would be a remarkable coincidence if they were the same today, but different at other times.  The E/M light speed obeys the relation
\begin{equation} 
c^2 = \frac{1}{{\epsilon_0 \mu_0}},
\end{equation} 
where $\epsilon_0$ and $\mu_0$ are the electric and magnetic ``constants'' of free space, resp.  If $c$ is variable, then either $\epsilon_0$ or $\mu_0$ or both must vary.

Current measurements with atomic clocks (\cite{Peik}, \cite{SGK}) have achieved an accuracy that indicate the frequency of atomic spectra do not change with time.  Of course, when measured on a frame moving at a different velocity or in a gravitational field, frequency does change.  There are also astronomical indications of a variation in $\alpha_F$ \cite{Webb}, but these are much smaller than would occur if $c(t)$ changed as calculated here.  On an inertial frame, this means that the fine structure constant  $\alpha_f$ and the Rydberg constant $R_\infty c$ (expressed as a frequency) do not change with $c(t)$.

The fine structure constant $\alpha_f$ in SI units \cite{C} is
\begin{equation}
	\alpha _f = \frac{e^2}{4\pi \epsilon_0 \hbar c},
\lbl{d9}
\end{equation}
and the Rydberg frequency is
\begin{equation}
	R_\infty c = \alpha _f ^2 \frac{m_e c^2}{4 \pi \hbar} = \frac{e^4 m_e}{\epsilon_0 ^2(4\pi \hbar)^3}.
\lbl{d10}
\end{equation}
Because $\alpha_F$ is dimensionless, the $4\pi \epsilon_0$ is often omitted in the fine structure constant since it is unity in Gaussian coordinates, but it is essential here if we are to consider a variable $c(t)$ for the universe.  

For these to remain constant while keeping $e$, $\hbar$ and $mc^2$ constant requires that
\begin{equation}
	\epsilon_0(t) c(t) = \frac{1}{\mu_0(t) c(t)} \equiv k = \epsilon (t_0) c (t_0), \; \mbox{a constant}.
\lbl{d11}\end{equation}
$ \frac{1}{k} = \sqrt{\frac{\mu  _0}{\epsilon _0}} $, the impedance of free space.
This assumption means that the electrostatic repulsion $f$ between two electrons will vary:
\begin{equation} 
f = -\frac {e^2}{\epsilon_0 R^2}.
\lbl{d9'}\end{equation} 

Maxwell's equations in 3D field vectors in the rest frame of FLRW with a constant speed of light \cite{Str} are 
\begin{equation} \begin{array}{rcl}
	\mbox {curl} E + \frac{\partial B}{\partial t} = 0, \\[.15cm]	
	\mbox {curl} H - \frac{\partial D}{\partial t} = J, \\[.15cm]
	\mbox {div} D = \sigma, \\[,15cm]
	\mbox {div} B = 0, \\[,15cm]
	\mbox {div} J + \frac{\partial \sigma}{\partial t} = 0.
\end{array} \lbl{d12}
\end{equation}
Scaler $\phi$ and vector $A$ potentials can be introduced such that
\begin{equation} \begin{array}{rcl}
	B = \mbox {curl} A, \\[.15cm]
	E = -\mbox{grad} \phi - \frac{\partial A}{\partial t},
\end{array} \lbl{d13} \end{equation}
and the equation for the force on a particle with charge $q$, mass $m$, and velocity $V$ is (see \cite[p118]{brg})
\begin{equation}
	m\frac{d (\gamma V)}{dt} = q(E + V \otimes B)
\lbl{d14} \end{equation}

With the use of $\hat{t}$ and eq \ref{d11} and with the relations $D = \epsilon _0 E, B = \mu_0 H$ for free space, these can be converted 
to exactly the same equations by replacing $t$ by $\hat{t}$ and by replacing the field variables by hat variables so that the partial time derivatives of hat variables do not include $\epsilon_0, \mu_0$, or $c$ except in combinations equaling $k$, a constant.  This is accomplished by the following: $\b = k B = \hat{H} = \frac{H}{c}$, $\hat{D} = D = \hat{E} = \epsilon_0 E$, $\hat{\sigma} = \sigma$, $\hat{J} = \frac{J}{c}$, $\hat{A} = k A$, $\hat{\phi} = \frac{c}{k} \phi$, and $\hat{q} = q$.  Thus, with hat variables and $\hat{t}$, Maxwell's equations have only two fields $\hat{E}, \hat{H}$ with no varying coefficients.  
\begin{equation} \begin{array}{rcl}
	\mbox {curl} \hat{E} + \frac{\partial \hat{H}}{\partial \hat{t}} = 0, \\[.15cm]	
	\mbox {curl} \hat{H} - \frac{\partial \hat{E}}{\partial \hat{t}} = \hat{J}, \\[.15cm]
	\mbox {div} \hat{E} = \hat{\sigma}, \\[,15cm]
	\mbox {div} \hat{H} = 0, \\[,15cm]
	\mbox {div} \hat{J} + \frac{\partial \hat{\sigma}}{\partial \hat{t}} = 0.
\end{array} \lbl{d12}
\end{equation}
The potential equations become
\begin{equation} \begin{array}{rcl}
	\hat{H} = \mbox {curl} \hat{A}, \\[.15cm]
	\hat{E} = -\mbox{grad} \hat{\phi} - \frac{\partial \hat{A}}{\partial \hat{t}},
\end{array} \lbl{d13} \end{equation}
Since they have no coefficients that vary with time, they are $\hat{L}$ covariant to frames with $d\hat{T} = c(t)dT$ replacing $d\hat{t}$ just like the original Maxwell's equations.  Thus, they are valid in every moving frame whose physical time is $T$.

With $\hat{V} = \frac{V}{c}$ and $\hat{m} = mc^2$ the pondermotive equation \ref{d14} becomes
\begin{equation}
	\hat{m} \frac{d(\gamma {\hat{V}})}{d\hat{T}} = \frac{q}{\epsilon _0}(\hat{E} + \hat{V} \otimes \hat{H}).
\lbl{6} \end{equation}
These all become the usual expressions when the speed of light is constant $c = 1$.  

E/M world vectors and tensors can be constructed in the usual way\cite{brg}.  
Thus, the extended potential vector is $\hat{\phi} _\mu = (\hat{A}_i ,-\hat{\phi})$, the extended charge  vector $\hat{\Gamma} ^\mu = (\hat{J}^i ,-\hat{\sigma})$, and the extended E/M field tensor is
\[
	\hat{F}_{\mu \nu} =
	\begin{pmatrix}
		& 0
		& -\hat{H}_3
		& +\hat{H}_2
		& -\hat{E}_1 \\[10pt]
		& +\hat{H}_3
		& 0
		& -\hat{H}_1
		& -\hat{E}_2 \\[10pt]	
		& -\hat{H}_2
		& +\hat{H}_1
		& 0
		& -\hat{E}_3 \\[10pt]
		& +\hat{E}_1
		& +\hat{E}_2
		& +\hat{E}_3
		& 0
\end{pmatrix}.\]

The field tensor can be obtained from the curl of the potential vector 
\begin{equation}
	\hat{F}_{\mu \nu} = \hat{\phi}_{\mu,\nu} -\hat{ \phi}_{\nu,\mu} \ ,
\lbl{sr7} \end{equation}
and Maxwell's equations become the divergence of the field tensor equaling the charge vector \cite[p113]{brg}:
\begin{equation}
	\hat{F}^{\mu \nu}_{\quad,\nu} = -\hat{\Gamma}^\mu.
\lbl{sr8} \end{equation}
The pondermotive equation for a particle of charge $q$ and mass $m$ becomes a force vector equaling $\hat{m}$ times an acceleration vector:
\begin{equation}
	\frac {q}{\epsilon_0} \hat{F}_{\mu \nu} \hat{U}^{\nu} = -\hat{m} \eta_{\mu \nu} \frac{d\hat{U}^\nu}{ds}. 
\lbl{sr9} \end{equation}
The SR stress-energy tensor is
\begin{equation} 
T^{\mu\nu} = \frac {1}{\epsilon_0}[F^\mu_\lambda F^{\nu \lambda} -\frac {1}{4}\eta^{\mu \nu}F^{\mu \sigma} F_{\sigma \nu}]
\end{equation} 
For GR with curvelinear coordinates, the stress-energy tensor is
\begin{equation} 
T^{\mu\nu} = \frac {1}{\epsilon_0}[F^\mu_\lambda F^{\nu \lambda} -\frac {1}{4}g^{\mu \nu}F^{\mu \sigma} F_{\sigma \nu}]
\end{equation} 
The dimensions of $\hat{E}$,$\hat{H}$, and $F^{\mu \nu}$ are electric charge per unit area, whereas for $T^{\mu \nu}$ it is energy per unit volume.  Because of $\epsilon_0$, both the force and the energy density are dependent on $c(t)$ just like the force between two electrons (eq \ref{d9'}).\\\

\subsection{The extended FLRW metric for a homogeneous and isotropic universe} \lbl{FLRW}

We assume that the concentrated lumps of matter, like stars and galaxies, can be averaged to the extent that the universe matter can be considered continuous, and that the surroundings of every point in space can be assumed isotropic and the same for every point.

By embedding a maximally symmetric (i.e., isotropic and homogeneous) three dimensional sphere, with space dimensions $r$, $\theta$, and $\phi$, in a four dimension space which includes time $t$, one can obtain a differential line element $ds$ \cite[page 403]{We1} such that
\begin{equation}
	ds^2 = g(t)dt^2 - f(t) \left[ \frac {dr^2}{1-kr^2} +r^2 d\theta^2 +r^2 sin^2\theta d \phi ^2 \right],
\end{equation}
where 
\begin{equation}
	r = \left \{ \begin{array}{l} \sin{\chi},~~~k = 1,\\ \chi,~~~~k = 0,\\ \sinh{\chi},~~~k = -1,
	\end{array} \right.
\lbl{2"} 
\end{equation}
$k$ is a spatial curvature determinant to indicate a closed, flat, or open universe, resp., and 
\begin{equation}
d\chi^2 \equiv \frac{dr^2}{1-kr^2}.
\end{equation}

We let $a(t) \equiv \sqrt{f(t)}$ be the cosmic scale factor multiplying the three dimensional spatial sphere, so that the differential radial distance is $a(t)d\chi$.

The $g(t)$ has normally been taken as $g(t) = c^2 = \mbox{constant}$, so that $c$ is the constant physical light speed and $t$ is the physical time on each co-moving point of the embedded sphere.  In both cases by physical, we mean that their value can represent measurements by physical means like standard clocks and rulers, or their technological equivalents.  In order to accommodate the possibility of the light speed being a function of time, we make $g(t) = c(t)^2$.  The resulting equation for the differential line element becomes an extended FLRW metric: 
\begin{equation}
	ds^2 = c(t)^2dt^2 - a(t)^2[d\chi^2 + r^2d\omega^2], 
\lbl{1}
\end{equation} 
where $d\omega^2 \equiv d\theta^2 + sin^{2} \theta d\phi^2 $.   
For radial world lines this metric becomes  Minkowski in form
with a differential of physical radius of $a(t)d\chi$.

It will be convenient to introduce the time related quantity $\hat{t}$, which we will call a generalized cosmic time, defined by
\begin{align}
\hat{t}  & \equiv \int_0^{t}{ c(t)dt},\\	t & = \int_0^{\hat{t}}{\frac{ d\hat{t}}{\hat{c}(\hat{t})}},
\lbl{3'}
\end{align}
where $\hat{c}(\hat{t}) = c(t)$, and where the lower limit is arbitrarily chosen as $0$.

The line element then becomes
\begin{equation}
	ds^2 = d\hat{t}^2 - a^2(d\chi^2 + r^2d\omega^2).
\lbl{3"}
\end{equation}

It should be emphasized that $\hat{t}$ itself is a legitimate more general coordinate.  $\hat{t}$ plays the same role in the FLRW space with a variable $c(t)$ as does $t$ for a FLRW space with constant $c$.
The physical time $t$ is a transform from it. $\hat{t}$ and its transform to $t$ allows for the physics to apply to a variable light speed.  

\subsection{Unchanged GR field equation for c(t)} \lbl{FE}

We assume the standard action of GR without any non-standard additions that some have used to produce the variable light speed\cite{Mag3}.  We allow the metric that determines the curvature tensor to introduce the varying light speed.  This will create a relationship between the varying light speed and the components of the stress-energy tensor.  In addition we use the Lagrangian $L_{se}$ of the extended stress-energy tensor. In order to use the standard GR action, we assume that $\frac{G}{c^4} \equiv \hat{G} = \frac{G_0}{c_0^4}$ is constant.  This is needed to keep constant the Newtonian energy $-\frac{Gm_1 m_2}{R}$ when the rest energy of mass $mc^2$ is constant.  We also assume that $\Lambda$ is  constant, possibly representing some kind of vacuum energy density.  :
\begin{equation}  
S = \int \sqrt{-g}(R - 2\Lambda + 16\pi \hat{G} L_{se})d^4 \xi.
\end{equation} 
$R$ is the Ricci scalar for the metric 
 \begin{equation}
	ds^2 = g_{\mu \nu}d\xi^\mu d\xi^\nu,
 \end{equation}
and $g$ is the determinate of $g_{\mu \nu}$.
  Minimizing the variation of $S$ with $g_{\mu \nu}$, we get the usual GR field equation:
\begin{equation}
	G_{\mu \nu} + \Lambda g_{\mu \nu} = 8\pi \hat{G} T_{\mu \nu}.
\lbl{d5}
\end{equation}
$G_{\mu \nu}$ is calculated for a particular metric using the usual Riemannian geometry.

\subsection{GR for FLRW universe with c(t)} \lbl{FE1}

We will now apply this field equation to an ideal fluid of density $\rho$ and pressure 
$p$ in a homogeneous and isotropic universe for which the extended FLRW line element 
in the variables $t,r,\theta,\phi$ is (eq \ref{1}):
\begin{equation}
	ds^2 = c(t)^2dt^2 - a^2(\frac {dr^2}{1-kr^2} + r^2d\theta^2 + r^2cos^2\theta d \phi^2).
\lbl{d5'}
\end{equation}

 As $ds$ is written in eq \ref{d5'}, the components of $G_{\mu \nu}$ will contain first and second derivatives of $c(t)$.
In order to find a solution to the field equation we will transform the time variable $t$ to $\xi^4 = \hat{t}$.  This will not change the relation of $G_{\mu \nu}$ to $T_{\mu \nu}$, but will eliminate the derivatives of $c(t)$ in $G_{\mu \nu}$ and transform $G_{\mu \nu}$ to a known solution.   $t$ is still the observable, and $\hat{t}$ is a non-physical transform from it.
For a perfect fluid of pressure $p$ and mass density $\rho$, we can define $\hat{\rho}\equiv \rho c^2$ so that $\hat{\rho}$ obeys the same conservation and acceleration laws using $d\hat{t}$ as does $\rho$ using $dt$ (Sect \ref{gr}).  
We can then write the two significant field equations \cite[page 729]{MTW} for $a(\hat{t})$ as
\begin{equation}
	\frac{3 \dot{a}^2}{a^2} + \frac{3k}{a^2} - \Lambda = 8 \pi \hat{G} \hat{\rho} ,
\lbl{d6}
\end{equation}
and
\begin{equation}
	+2 \frac{\ddot{a}}{a} +\frac{\dot{a}^2}{a^2} +\frac{k}{a^2} -\Lambda = -8 \pi \hat{G} p,
\lbl{d7}
\end{equation}
where the dots represent derivatives with respect to $\hat{t}$.  All variables (including $\hat{t}$ and $a$) are in standard units. Eq \ref{d6} can be solved to give $a$ as a function of $\hat{t}. \rho, k$, and $\Lambda$.  When we know 
$c(\hat{t})$, we can obtain the observables $a(t)$ and $c(t)$ by transforming $\hat{t}$ back to $t$. Solutions of these equations are carried out in Sect \ref{c} for a particular $c(t)$.  

We would now like to show that the proposed variation of $G(c)$ and $m(c)$ is internally consistent.   
Eq \ref{d6} with $\hat{G} = \frac{G}{c^4}$ and $\hat{\rho} = \rho c^2$ can be multiplied by $\frac{a^3}{3}$, differentiated, and subtracted from $ \dot{a} a^2$ times eq \ref{d7} to give
\begin{equation}
	\frac{d}{d \hat{t}}(\frac{G \rho a^3}{c^2}  ) = -\frac{3G}{c^4}\dot{a} a^2 p   ,
\lbl{d8}
\end{equation}
For small $p$, 
\begin{equation}
	\frac {G \rho a^3}{c^2} = \mbox{constant}
\lbl{d8'}
\end{equation}
If the energy density consists of $n$ particles per unit volume of mass $m$, so $\rho = nm$, then the conservation of particles requires $na^3$ be constant (for small velocities).  This makes 
\begin{equation}
	\frac{Gm}{c^2} = \mbox{constant}.
\lbl{d8"}
\end{equation}
This is consistent with our assumptions that $\frac{G}{c^4}$ and $mc^2$ are constant.

\section*{Acknowledgments} \lbl{ack}
\addcontentsline{toc}{section}{Acknowledgments}

I wish to acknowledge the invaluable help given by Paul Fife, University of Utah Mathematics Dept. 
I also wish to thank David W. Bennett, Univ. of Utah Philosophy Dept, for his support and
encouragement and the faculty and facilities of the Physics and
Astronomy Department of Tufts University for the number of years
that I was allowed to visit there.  I thank Richard Price, Univ. of Utah Physics Dept,
for valuable discussions during the earlier part of this investigation, and Ramez Atiya for insightful discussions of earlier drafts of this paper.  I would also thank my family for the considerable help in editing.  However, I alone am responsible for any errors of mathematics or interpretation that may be here.


\section*{Figures} 
\addcontentsline{toc}{section}{Figures}

\begin{figure}[tbp] 
  \centering
  \includegraphics[bb=8 8 428 320,width=5.67in,height=4.21in,keepaspectratio]{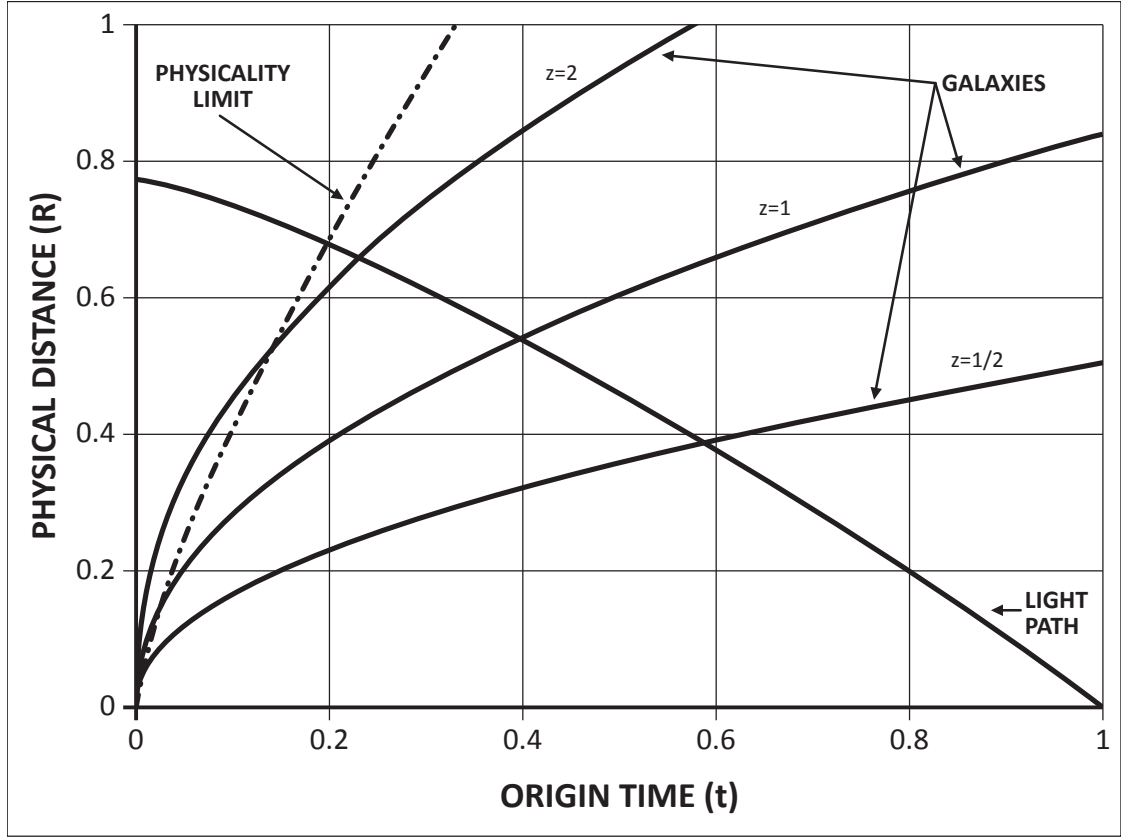}
  \caption{Physical distances ($\frac{R}{c_0t_0}$) for $\Omega = 1 $
 plotted against the normalized time on clocks at the origin ($\frac{t}{t_0}$) for various galaxy paths (labeled by their red shift $z$) and for the light path which the photons take after emission by any galaxy that arrives at the origin at $\frac{t}{t_0} = 1$.   Notice  that the slope of this light path close to the origin is $c_0 = 1$, where $V^2 << c^2$.  The light path starts at the far horizon at $t=0$, traveling monitonically towards the origin, but slower than its present speed in these non-local coordinates (like the Schwarzschild coordinates).  The galactic paths show the expanding universe in physical coordinates, some traveling faster than the light speed in these non-local coordinates.  
The dotted line shows the approximate upper limit of physicality, where both $R$ and the transformed time $T$ are physical.}
  \label{fig:c_t_Fig1}
\end{figure}

\begin{figure}[tbp] 
  \centering
  \includegraphics[bb=8 8 428 320,width=5.67in,height=4.21in,keepaspectratio]{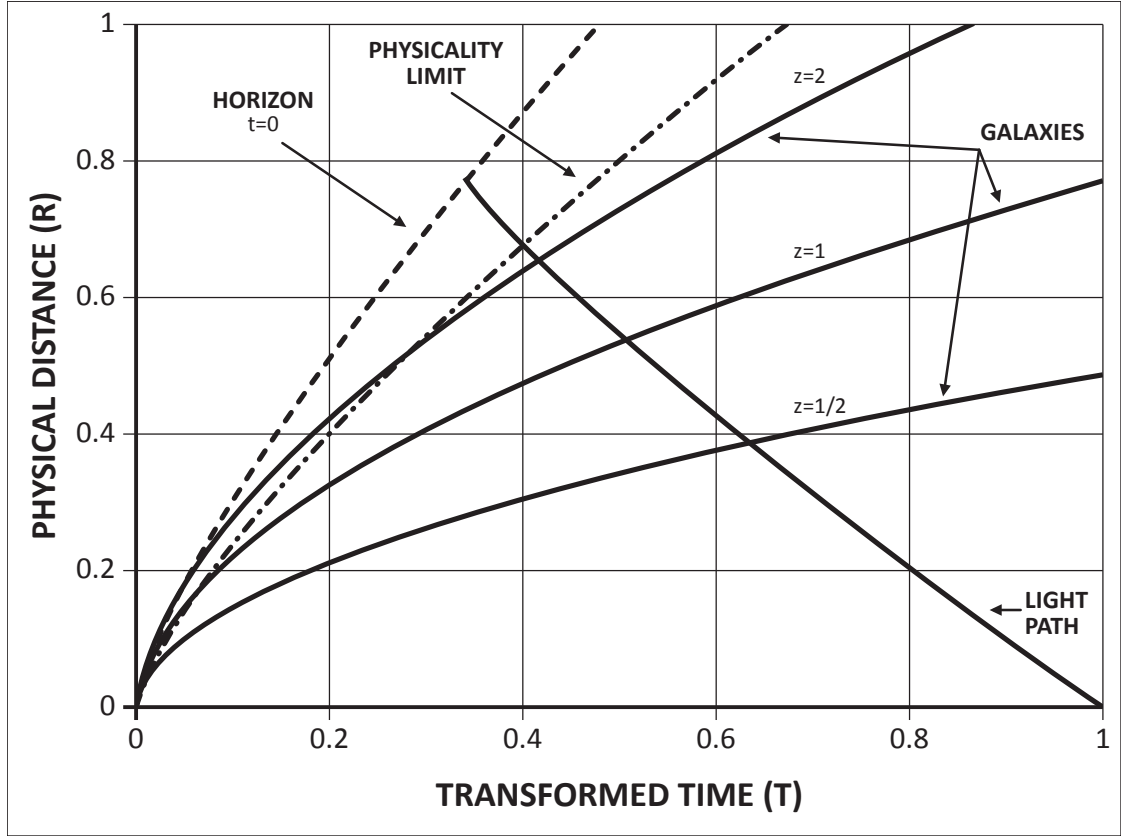}
  \caption{Physical distances ($\frac{R}{c_0t_0}$) for $\Omega = 1 $ vs the transformed time 
($\frac{T}{t_0}$) on clocks attached at $R$ for various galaxy paths (labeled by their red shift 
$z$) and for the light path that arrives at the origin at $T = t_0 = .75(\frac{1}{c_0H'_0})$.  The horizon is the locus of points where $t = 0, \gamma = \infty$.  The heavy dotted line shows the approximate upper limit of physicality for the transformed coordinates  ($A = .95$, that is $<5\%$ error in physical time rate $T_t$).  The slope of the light path is very close to $c(t)$ out to the limit of physicality. Light is emitted at finite $T$ allowing transformed time for galactic points to move out from $R=0$ before emitting their light we can see.}
  \label{fig:c_t_Fig2}
\end{figure}

\begin{figure}[tbp] 
  \centering
  \includegraphics[bb=8 8 428 320,width=5.67in,height=4.21in,keepaspectratio]{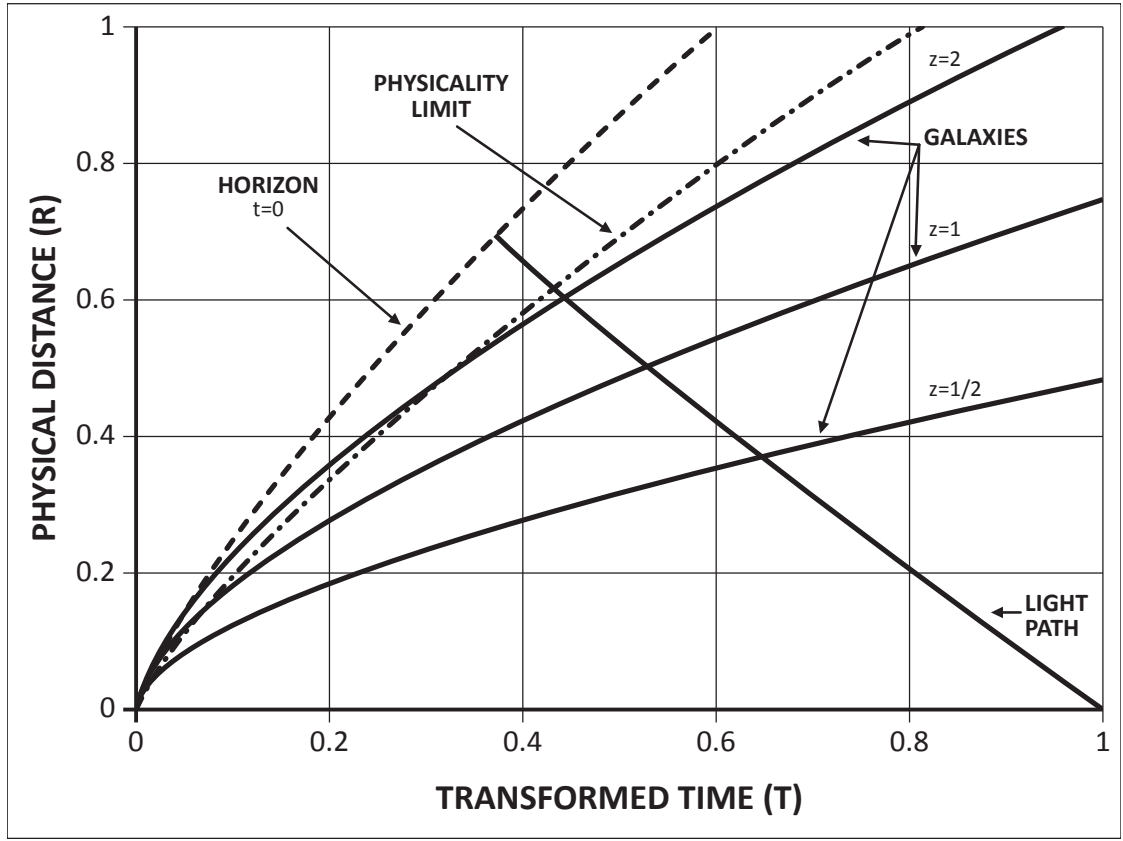}
  \caption{Physical distance($\frac{R}{c_0t_0}$) for lower density universe ($\Omega = \frac{1}{2}, \Omega _r = \frac{1}{2}) $  plotted against the transformed time ($\frac{T}{t_0}$) on clocks attached at $R$ for various galaxy paths (labeled by their red shift $z$) and for the light path that arrives at the origin at $T = t_0 = .767(\frac{1}{c_0H'_0})$. The horizon ($t = 0$) and the physicality line ($A = .96$) occur at later times and shorter distances than for a flat universe (Fig 2), but not as much as for the empty universe (Fig 5).  Similarly, the light path is straighter than Fig 2, but not as straight as Fig 5.}
  \label{fig:c_t_Fig3}
\end{figure}

\begin{figure}[tbp] 
  \centering
  \includegraphics[bb=8 8 507 320,width=5.67in,height=3.54in,keepaspectratio]{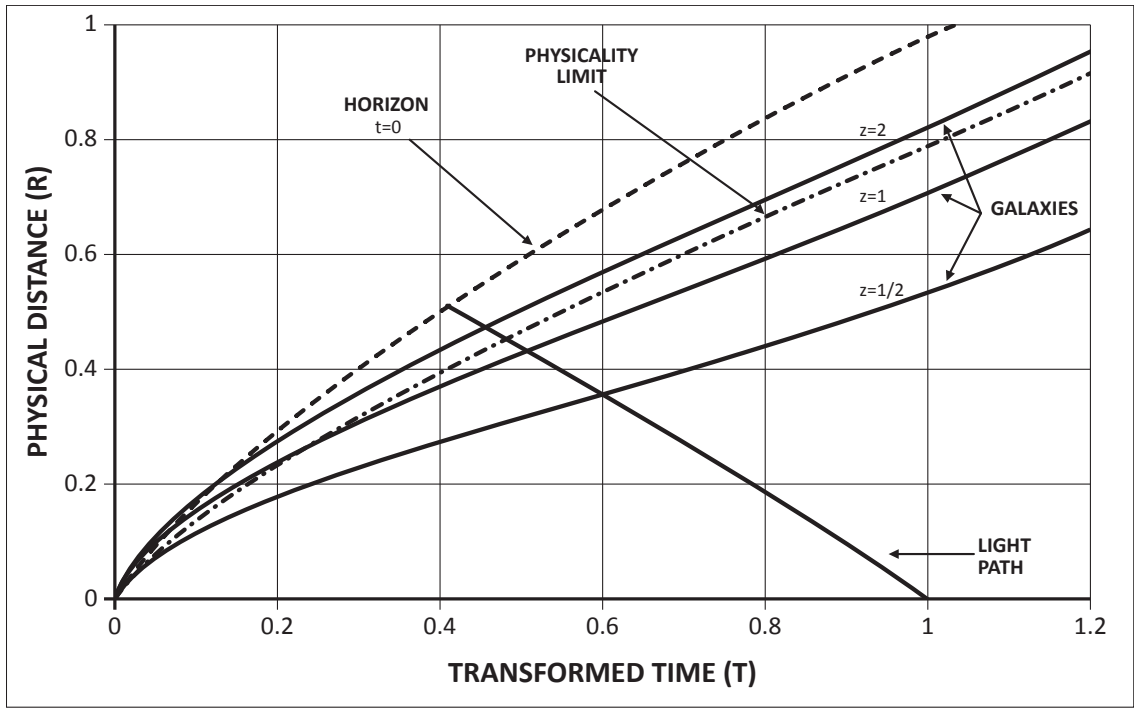}
  \caption{Physical distance ($\frac{R}{c_0 t_0}$) vs. transformed time ($\frac{T}{t_0}$) for dark energy ($\Omega = \frac{1}{4}, \Omega _\Lambda = \frac{3}{4}$).  The present time $t_0 = .407(\frac{1}{c_0H'_0})$.  Notice the inflection points on all curves where the dark energy density becomes larger than the matter density.}
  \label{fig:c_t_Fig4}
\end{figure}

\begin{figure}[tbp] 
  \centering
  \includegraphics[bb=8 8 428 320,width=5.67in,height=4.21in,keepaspectratio]{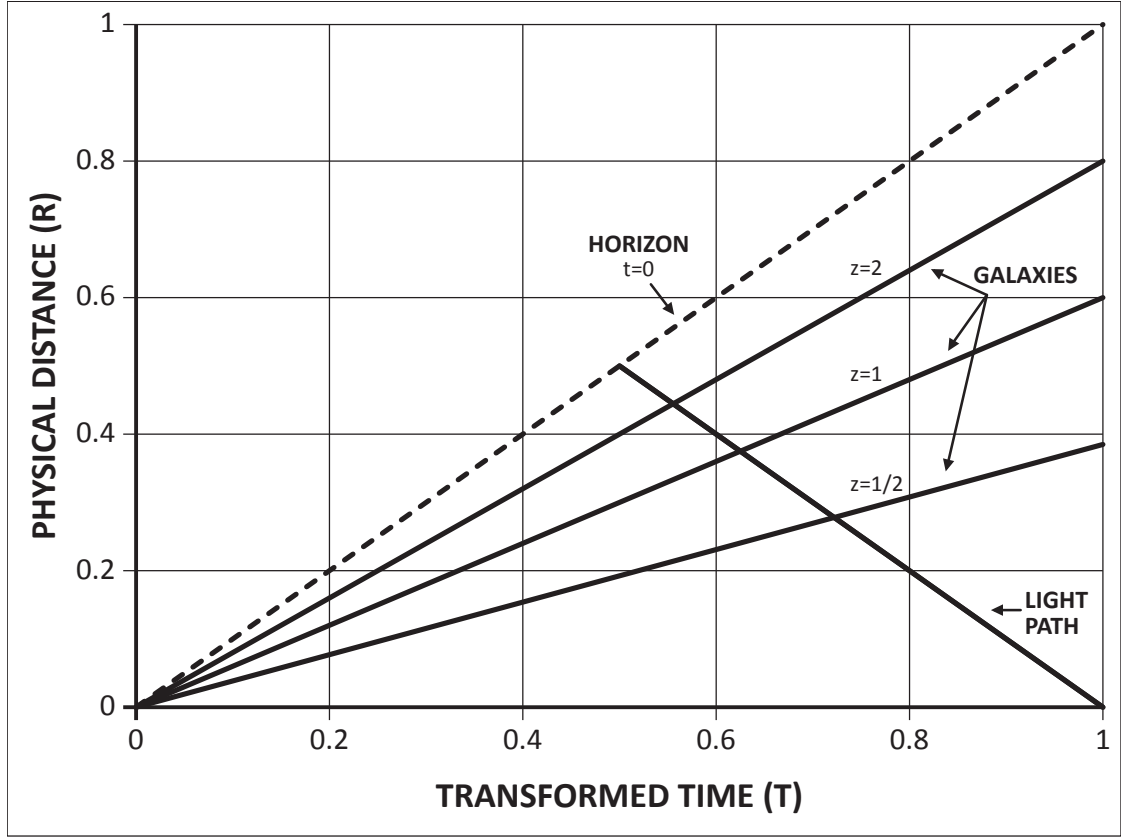}
  \caption{Physical distance($\frac{R}{c_0t_0}$) for the empty expanding universe ($\Omega = 0, \Omega _r = 1) $  plotted against the transformed time ($\frac{T}{t_0}$) on clocks attached at $R$ for various galaxy paths (labeled by their red shift $z$) and for the light path that arrives at the origin at $T = t_0$.  The horizon is the locus of points where $t = 0$.  All lines are straight and physical, since there is no space curvature, and the light speed is $c(t) = c_0$.  The remotest galactic point travels from the origin at $T = 0$ out to $\frac{c_0t_0}{2} $ at the light speed $c_0$.}
  \label{fig:c_t_Fig5}
\end{figure}

\end{document}